\documentclass[a4paper,11pt]{article}
\usepackage{multirow}
\usepackage{jinstpub} 
\usepackage{gensymb}
\usepackage{float}
\usepackage{subcaption}
\usepackage{graphicx}
\usepackage{placeins}
\usepackage[table,xcdraw]{xcolor}

\usepackage{lineno}
\title{\boldmath Test beam measurement of ATLAS ITk Short Strip module at warm and cold operational temperature}

\author[a]{J.-H.~Arling,}
\author[a]{C.~Becot,}
\author[b]{E.~Buchanan,}
\author[c]{J.~Dopke,}
\author[c]{B.~Gallop,}
\author[d]{J.~Kaplon,}
\author[e]{J.~S.~Keller,}
\author[f]{J.~Kroll,}
\author[b]{Y.~Li,}
\author[g,1]{Z.~Li \note{Corresponding author.},}
\author[i]{J.~Liu,} 
\author[a]{Y.~Liu,} 
\author[e]{S.~Y.~Ng,} 
\author[h,f]{R.~Privara,}
\author[a]{A.~Renardi,}
\author[i]{A.~Rodriguez~Rodriguez,}
\author[a]{E.~Rossi,}
\author[i]{F.~Ruehr,}
\author[c]{C.~Sawyer,}
\author[i]{D.~Sperlich,}
\author[g]{A.~R.~Weidberg,}
\author[j]{D.~F.~Zhang}

\affiliation[a]{Deutsches Elektronen-Synchrotron DESY, Hamburg and Zeuthen, Germany}
\affiliation[b]{Beijing Institute of High Energy Physics, Chinese Academy of Science, Beijing, China}
\affiliation[c]{Particle Physics Department, Rutherford Appleton Laboratory, Didcot, United Kingdom}
\affiliation[d]{CERN, Geneva, Switzerland}
\affiliation[e]{Department of Physics, Carleton University, Ottawa ON, Canada}
\affiliation[f]{Institute of Physics of the Czech Academy of Sciences, Prague, Czech Republic}
\affiliation[g]{Department of Physics, Oxford University, Oxford, United Kingdom}
\affiliation[h]{Faculty of Science, Palacký University, Olomouc, Czech Republic }
\affiliation[i]{Physikalisches Institut, Albert-Ludwigs-Universität Freiburg, Freiburg, Germany}
\affiliation[j]{Physics Department, Tsinghua University, Beijing, China}

\emailAdd{zhiying.li@mansfield.ox.ac.uk}

\abstract{
This study is focused on an investigation of the performance of the Short Strip module developed by the ATLAS Inner Tracker (ITk) strip collaboration  using electron beams of energy \mbox{5.4 GeV} and  \mbox{5.8 GeV} at the DESY-II Testbeam Facility.  The noise at $+30 \:\degree\mathrm{C}$ and $-30 \:\degree\mathrm{C}$ was measured. The ratio of the two measurements is compared with a circuit-model calculation.  
The measured noise at $-30\:\degree\mathrm{C}$ is compared with the maximum noise that would correspond to an acceptable signal-to-noise ratio after the expected radiation damage from operation at HL-LHC.
The measured charge distributions at $+30 \:\degree\mathrm{C}$ and $-30 \:\degree\mathrm{C}$ are compared with GEANT4 simulations. 
The detection efficiency and noise-occupancy were measured as a function of threshold at both $+30 \:\degree\mathrm{C}$ and $-30 \:\degree\mathrm{C}$.
The average cluster width was measured as a function of threshold.
Scans of detection efficiency versus threshold at different delay settings were used to reconstruct the pulse shape in time. The resulting pulse shape was compared with a circuit model calculation.
}


\begin{document}
\maketitle
\flushbottom
\section{Introduction}
\par
Test beams enable controlled studies of the performance of tracking detectors. In the ATLAS Inner Tracker (ITk) upgrade they are used to validate the performance of devices, such as modules, designed for the High Luminosity Upgrade of the LHC (HL-LHC). This allows key performance parameters such as hit efficiency for charged particles to be studied in detail prior to the assembly of the final detector~\cite{collaboration2017technical} \cite{abcstarModule}.
\par
In ATLAS operation the ITk strip modules will be operated at cold temperatures \cite{collaboration2017technical}, it is therefore important to compare module performance at cold and warm temperatures. 
A brief review of the ATLAS tracker upgrade (ITk) is given in section~\ref{ATLAS-ITk}.
The experimental setup at the DESY test beam is described in section~\ref{Setup}.
The analysis of the test beam data is presented in section~\ref{analysis}. The ITk detector will need to be operated at temperatures down to $-35\:\degree\mathrm{C}$ to minimise radiation damage. However, the detector will be assembled at room temperature and initial testing will be performed in these conditions. Comparisons are made for the noise performance and signal charge measurements for cold and warm temperatures. The timing response of the module was partially measured and compared to calculations. 

\section{ATLAS ITk Upgrade}
\label{ATLAS-ITk}
\subsection{High Luminosity LHC and ATLAS Inner Tracker}
The Large Hadron Collider (LHC) installed at CERN is going to be upgraded to the High-Luminosity LHC (HL-LHC), with a nominal luminosity of $5\times10^{34}\;\mathrm{cm}^{-2}\mathrm{s}^{-1}$ and a peak luminosity of $7.5\times10^{34}\;\mathrm{cm}^{-2}\mathrm{s}^{-1}$. The integrated luminosity of HL-LHC will be an order of magnitude larger than that of LHC, thus greatly expanding its physics potential~\cite{collaboration2017technical}. The increase in luminosity requires upgrades of most of the detectors of the ATLAS experiment with the largest upgrade planned for the ATLAS inner detector, which will be replaced with the new all-silicon inner tracker (ITk), providing improved radiation tolerance, better granularity and faster detector readout.

\par
\begin{figure}
    \centering
    \includegraphics[width=0.8\textwidth]{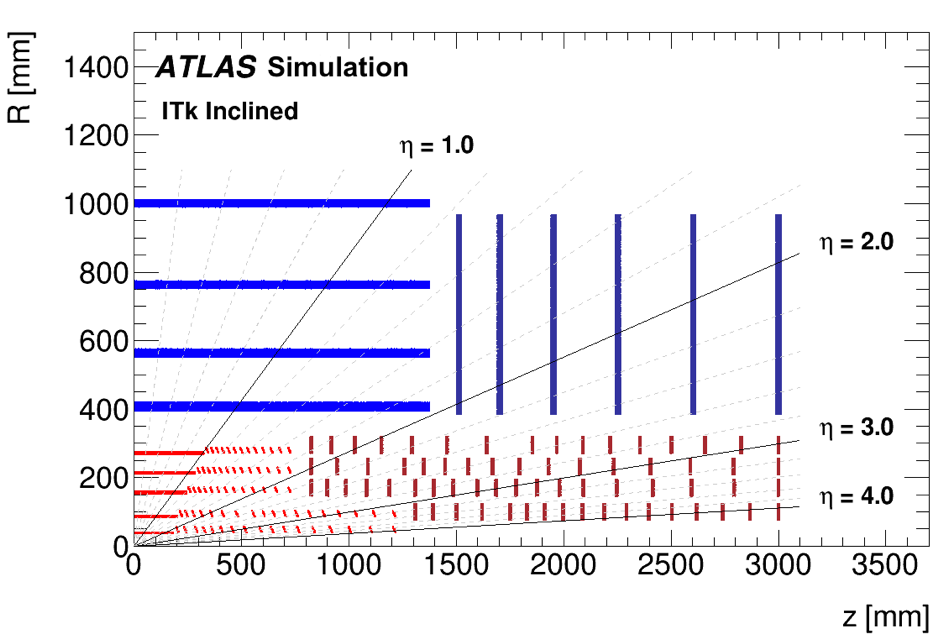}
    \caption{Schematic layout of one quadrant of the ATLAS ITk detector. The red color corresponds to the ITk pixel sub-detector, while the blue components represent the ITk strip sub-detector. The horizontal axis goes along the beam line and the origin of the axes lies in the interaction point. The vertical axis corresponds to the radius measured from the beam axis \cite{collaboration2017technical}.}
    \label{fig:plot10}
\end{figure}

\begin{figure}
    \centering
    \includegraphics[width=0.8\textwidth]{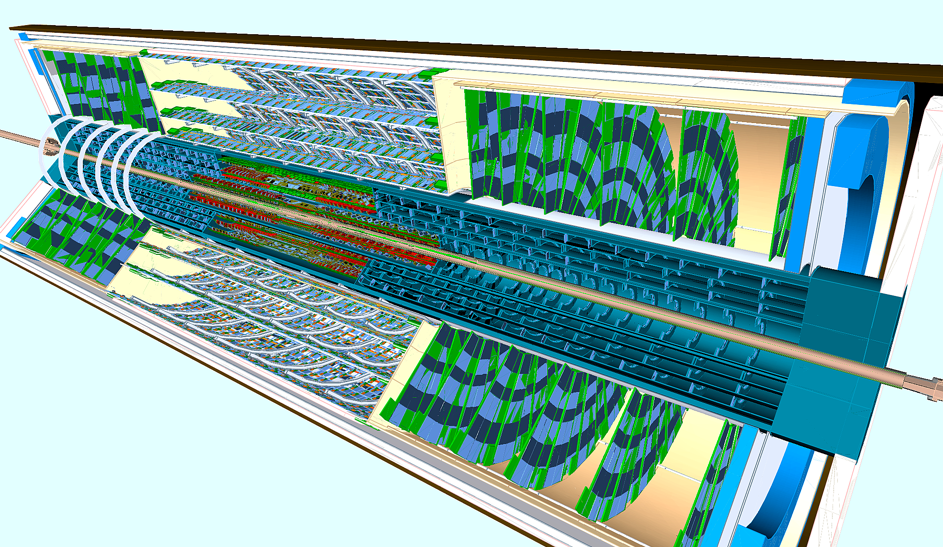}
    \caption{Simulated layout of the ATLAS ITk detector, with the inner pixel sub-detector surrounded by the strip sub-detector  \cite{collaboration2017technical}.}
    \label{fig:plot11}
\end{figure}

 Five concentric pixel barrel layers will be surrounded by 4 strip barrel layers. Both pixel and strip systems will include also their end-cap parts, consisting of the pixel rings of different radii and six strip disks on each side of the pixel and strip barrel, respectively. A cross-section of one quadrant of the ITk detector is shown in Figure~\ref{fig:plot10}, while
its simulated layout is illustrated  
in Figure~\ref{fig:plot11}.

The ATLAS ITk strip detector~\cite{collaboration2017technical}, composed of four concentric barrel layers and one end-cap section with six disks on each of its barrel sides, covers the pseudo-rapidity range up to $|\eta|=2.7$. The barrel layers are parallel to the beam line and range from \mbox{-1400 mm} to \mbox{+1400 mm} about the nominal collision point. Each barrel layer is built from staves, that provide mechanical, electrical and thermal support to the barrel modules. The inner two barrel layers accommodate the Short Strip modules (SS modules), having the sensors with parallel strips  of 24.16 mm length, while the outer two layers include Long Strip modules (LS modules) with 48.35 mm strips.  The pitch between two strips is 75.5 $\mu$m and the strip implants have a width of $16\:\mu$m.  The end-cap modules of 6 different kinds are mounted on petals, where 32 petals compose one end-cap disk. Strips of the end-cap sensors are approximately radially distributed and point to the beam axis centre, with strip lengths varying from \mbox{19.0 mm} to \mbox{60.1 mm}.  The end-cap sensor pitch ranges from 69 $\mu$m to 85 $\mu$m.
\par
\begin{figure}
    \centering
    \includegraphics[width=0.8\textwidth]{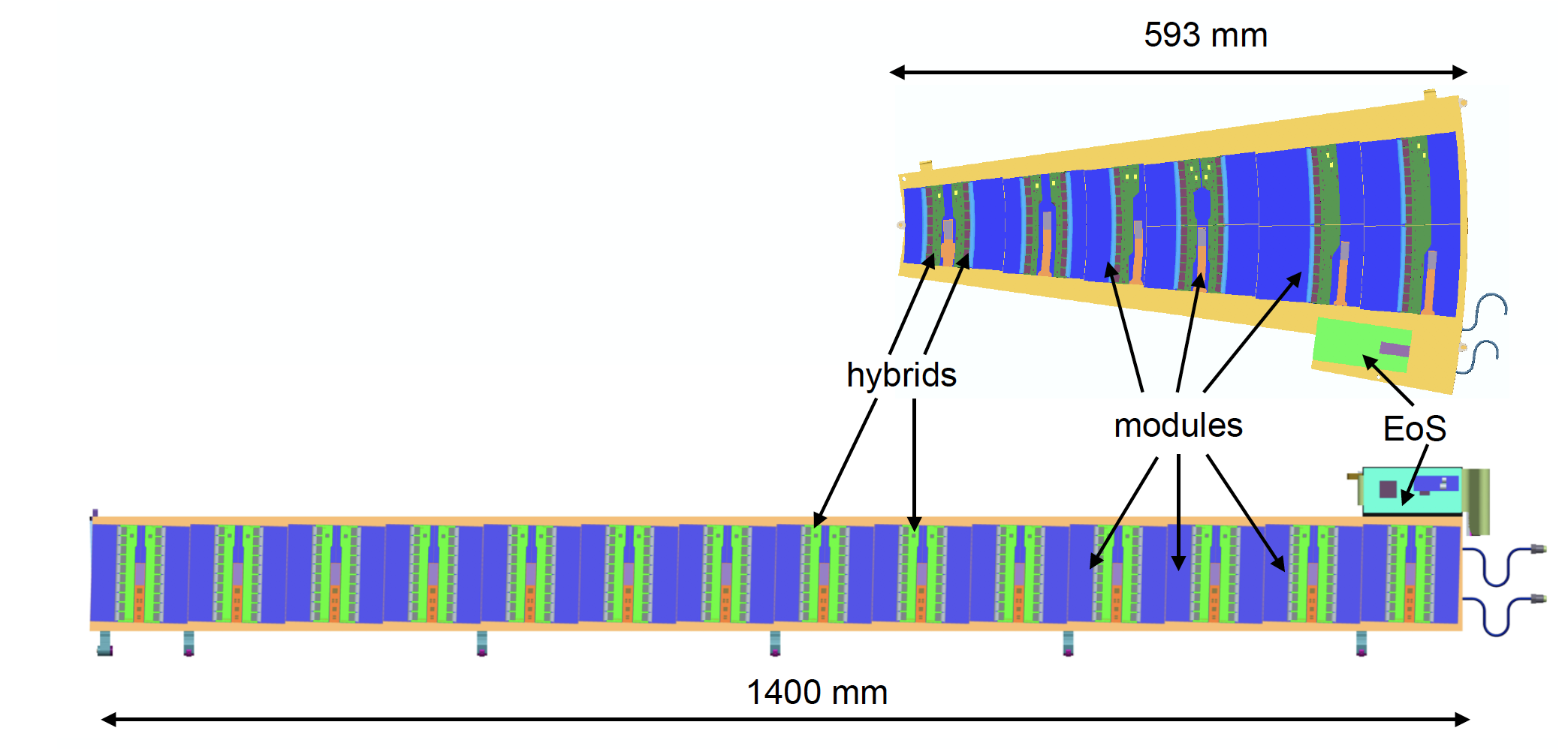}
    \caption{End-cap and barrel modules mounted on the petal (top) and stave (bottom), respectively. The end-of-stave (EoS) card provides the data and power connection between the modules and off-detector electronics. Optoelectronics is used for data transmission.}
    \label{fig:plot12}
\end{figure}

 A small stereo angle is imposed on the modules mounted on either side of the stave or petal that face each other.
This small stereo angle provides an enhanced resolution in Z (see Figure \ref{fig:plot10}) in the barrel modules and  radial measurement in the end-cap modules. In the barrel system, the modules on each side of the stave are rotated by $\pm26\;\mathrm{mrad}$ with respect to the beam line resulting in the total rotation angle of $52\;\mathrm{mrad}$ between the strips. In the end-cap system the total stereo angle of $40\;\mathrm{mrad}$ is achieved by the implementation of 20 mrad stereo angle directly in the strip sensor design. The stave and petal populated by the barrel and end-cap modules, respectively, are shown in Figure~\ref{fig:plot12}.

\begin{figure}
    \centering
    \includegraphics[width=0.8\textwidth]{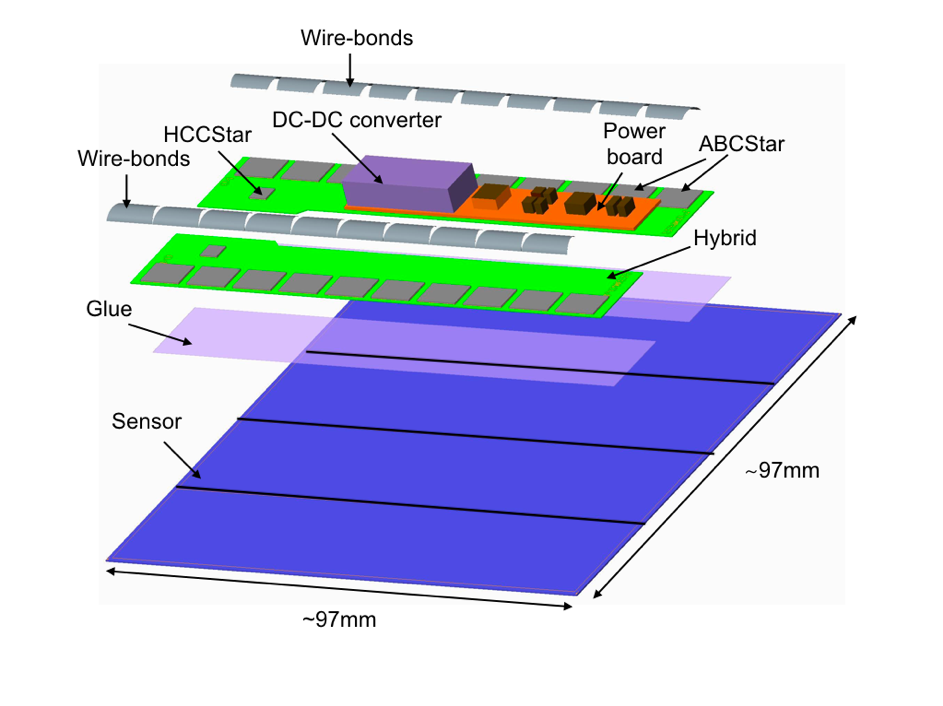}
    \caption{Components of the ITk Short Strip module \cite{collaboration2017technical}. The module has one sensor with two hybrids composed of ASIC chips and a power board glued on top. A short strip sensor is divided into 4 rows of strips as shown in the figure. The ASIC chips on each hybrid include one HCCStar chip and ten ABCStar chips. The HCCStar chip sends clock, command and reject signals to the ABCStar chips and data acquired from the ten ABCStar chips are multiplexed by the HCCStar chip to a signal output. The individual channels of ABCStar chips are connected to sensors via wire bonds.}
    \label{fig:plot13}
\end{figure}

\subsection{ATLAS ITk strip modules}
As shown in Figure~\ref{fig:plot13}, each ITk strip module consists of one silicon sensor, one or two hybrids and the powerboard. The silicon sensors include \mbox{n-type} implant strips on a \mbox{p-type} bulk with the active thickness of about $300\;\mu\mathrm{m}$. Using \mbox{n-in-p} sensors ensures that the detector signals arise mainly from electrons. Therefore, the transit time for the charge carriers is shorter than for \mbox{p-in-n} sensors because of the higher mobility of electrons compared to holes, hence the losses from charge trapping are reduced~\cite{unno2014development}. Additionally n-in-p sensors do not undergo type inversion with irradiation. The isolation between strips is realised with p-stop implants. The individual channels of the module readout are AC-coupled to the n-type strips.

The hybrid is a flexible PCB, on which the readout Application Specific Integrated Circuits (ASICs) are glued. For the module analysed in this paper, the relevant ASICs are the 256-channel ATLAS Binary Chips (ABCStarV0) \cite{ref2} and the Hybrid Controller Chips (HCCStarV0) \cite{collaboration2017technical} in their Star version. Ten ABCStar chips are connected to one HCCStar via direct communication ~\cite{abcstar}. Connection of the individual channels of ABCStar chips with the sensors strips is realized by wire bonds. 
A charged particle passing through the sensor creates a signal, which is transferred to the input channels of the ABCStar chips. 
The analog signal from each strip is then converted to binary hit/no-hit data. The front-end of the ABCStar chips features a charge sensitive pre-amplifier and gain stage, a shaper, a discriminator, a pipeline for 256 channels, and a buffer, as well as a cluster algorithm to compress data for output. 
The voltage signal after the gain stage is compared with a threshold in the discriminator to generate the binary output. The binary output is stored in the pipeline memory. On receipt of an L1 trigger, the binary data from hit strips are compressed into clusters and then transferred to the "L1" buffer before being read out~\cite{Pb}. The design ensures that the signal is contained within two LHC bunch crossings (\mbox{25 ns} each). In order to minimise pile-up in high-luminosity proton-proton collisions the 01X operation mode is used. In this mode, for data corresponding to a strip to be readout, the logic requires that there was a hit in the triggered bunch crossing and a no-hit in the preceding bunch crossing. There is no requirement on the presence of any hit in the succeeding bunch crossing. The ABCStar chip also has an internal circuit to inject a known amount of charge into the front end, which is used for calibration purposes.

The power board~\cite{Pb} provides a DC-DC converter, an additional ASIC called the Autonomous Monitoring and Control (AMACV2a) chip \cite{collaboration2017technical} and a GaNFET transistor. The DC-DC converter converts 11 V input to approximately 1.5 V required by the ASICs on the hybrid. The AMAC chip controls the low and high voltages and monitors the currents and temperature of the module. The transistor is a switch for the application of bias voltage on the sensor. The sensor bias voltage controlled by a GaNFET transistor should not significantly exceed the value of -500 V~\cite{abcstarModule}.

\section{Experimental Setup}
\label{Setup}
\subsection{DESY - Test Beam Area TB22}
\label{desybeam}
\par
\begin{figure}
    \centering
    \includegraphics[width=0.8\textwidth]{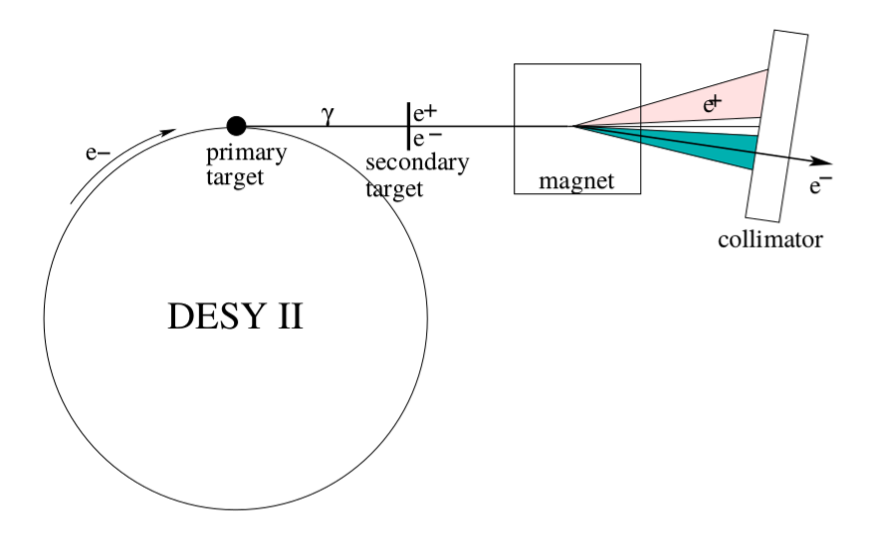}
    \caption{Electron generation for the test beam area 22 at DESY-II Testbeam Facility~\cite{DESY_plot}.}
    \label{fig:plot16}
\end{figure}

 Most of the analysis in this paper is based on data from the test beam campaign at the DESY-II Testbeam Facility that took place at DESY Hamburg in September 2019.  Data from the April 2019 campaign are also used for pulse shape profiling.  An electron beam is provided by the DESY-II electron synchrotron. The synchrotron has a circumference of 292.8 m and serves also as a pre-accelerator for the PETRA-III storage ring~\cite{DESY}. Carbon fibers can be moved into the beam to create bremsstrahlung photons, which leave the line tangentially.  The photons are then converted back to electrons with a secondary metal target. A dipole magnet is used to spread out the beam and select beam particles with a certain energy. The beam generation is illustrated in Figure~\ref{fig:plot16}. The maximum electron energy is 6 GeV. The electron beams used to obtain the results discussed in this paper was tuned to an energy of 5.4 GeV or 5.8 GeV.

\subsection{Device Under Test}
\begin{figure}
    \centering
    \includegraphics[width=0.8\textwidth]{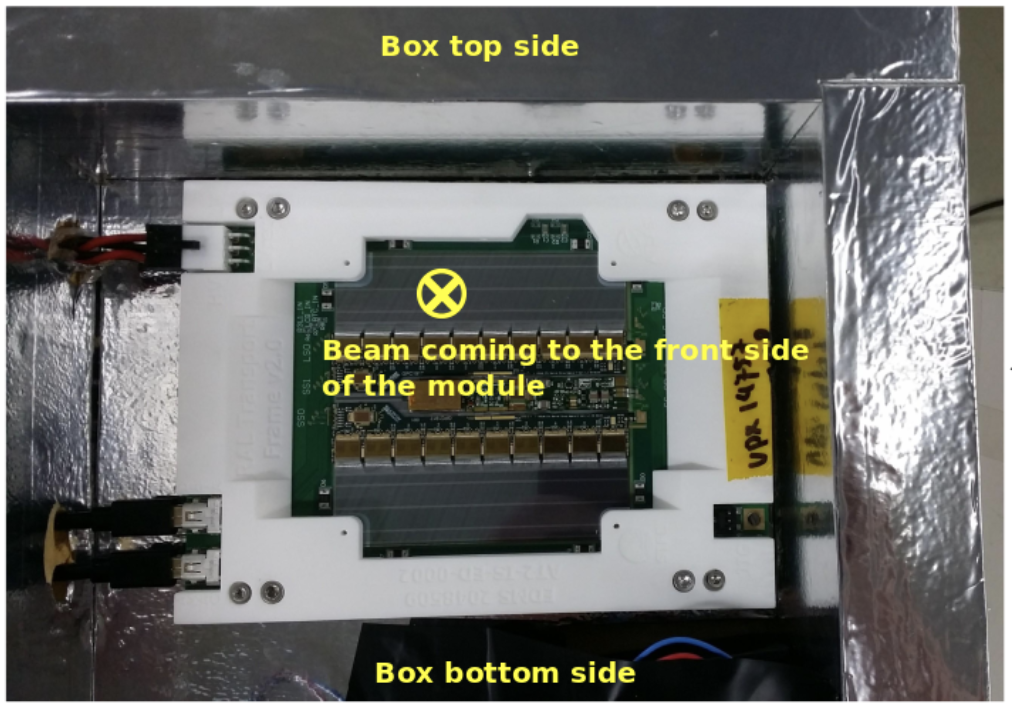}
    \caption{ITk SS module installed in the polystyrene cold box, which has been tested in September 2019. The beam was centred on chip 4 as indicated by the cross. The beam resulted in hits in approximately 150 strips. In this paper we have selected an area corresponding to a large number of incident electrons. This area covers 123 strips. }
    \label{fig:plot14}
\end{figure}

The device under test (DUT) measured during the test beam campaign in September 2019 was an unirradiated ITk SS barrel module as discussed in Sec. \ref{ATLAS-ITk} and it uses two hybrids as readout. 
Throughout the paper by "strip" we imply the region where a physical strip collects the signal. The bias voltage was \mbox{-400 V}, which is sufficient to ensure full depletion before irradiation. A photograph of the tested SS module installed in the polystyrene cold box is shown in Figure~\ref{fig:plot14}.

\subsection{The Beam Setup}
To measure the performance of the DUT, additional external detectors are needed in order to predict the track hit position on the DUT and to provide precise time matching between the DUT and external detectors. The DUT is positioned in an EUDET-type beam telescope system~\cite{eudet} composed of six Mimosa26 detectors~\cite{mimosa} with the Monolithic Active Pixel Sensors with binary readout, which provides independent reference tracking information. The sensor of the Mimosa26 detector has a thickness of $50\;\mu\mathrm{m}$ and consists of 576 by 1152 pixels with a pitch of $18.4\;\mu\mathrm{m}$, which results in an active area of $10.6\times21.2\;\mathrm{mm}$. The spatial resolution from the telescope system is much better than that of the DUT and typically reaches $5-10\;\mu\mathrm{m}$ at the DESY-II Testbeam Facility. While the trajectory of the test beam particles is precisely determined by the beam telescope, extra timing information is necessary for the measurement of the detection efficiency. The reason for this is the much longer readout window of the telescope compared to the DUT designed for the LHC cycle. Hence a timestamping of tracks to match the hits to the correct bunch crossing is required for the efficiency measurements. 
This is provided by the FE-I4 pixel detector with the USBPix readout~\cite{FEI4}.

\begin{figure}
    \centering
    \includegraphics[width=0.8\textwidth]{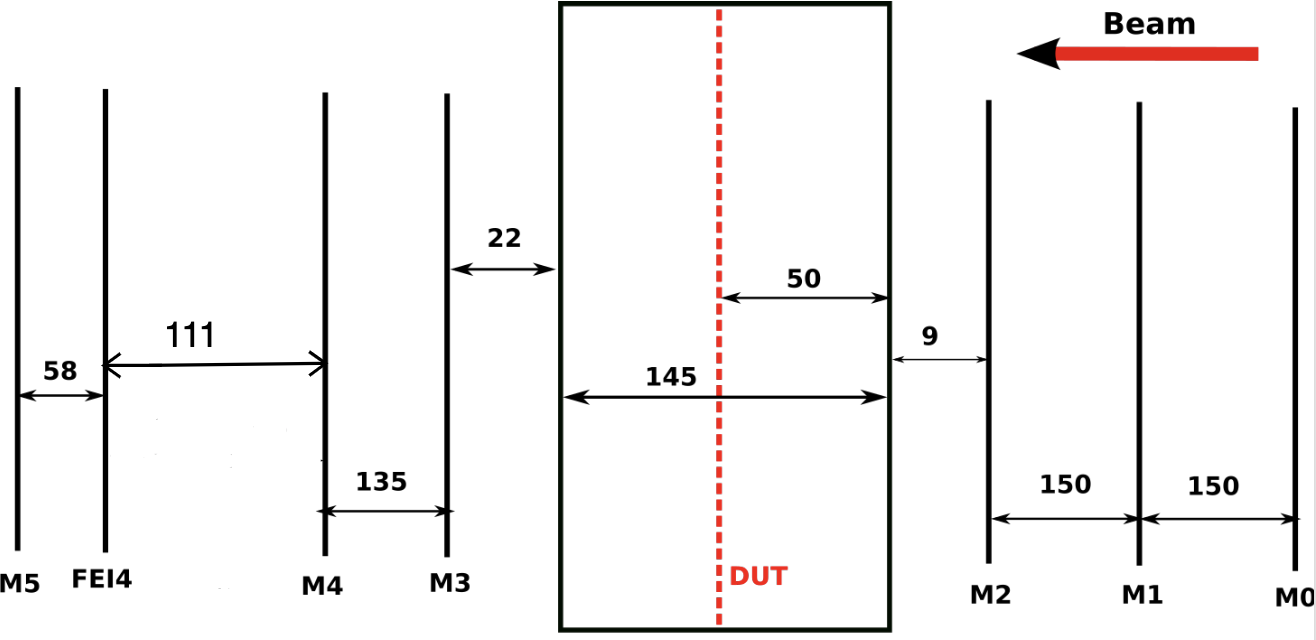}
    \caption{Geometry of the experimental setup used for testing of the ITk SS module in September 2019. The EUDET telescope planes with Mimosa26 pixel detectors are labeled from M0 to M5. The numbers are the distances between planes in mm.
    In addition there are scintillator planes upstream and downstream of the DUT, which provide external triggers to the DUT.}
    \label{fig:plot15}
\end{figure}

The geometry of the experimental setup is illustrated by the scheme in Figure~\ref{fig:plot15}, where the individual Mimosa26 planes have labels M0 to M5, with the index increasing in the direction of the beam. The rectangle around the DUT indicates a cooling box, in which the DUT is placed. The cooling box was cooled with dry ice to $-30\:\degree\mathrm{C}$.  The temperature is measured via a thermometer (thermocouple type K connected to a multimeter) located close to the module inside the cooling box. The cooling box with the module is positioned in the EUDET telescope facing the beam, with three Mimosa planes in front and behind the cooling box.

\subsection{Data Reconstruction}

The raw data taken during the test beam are processed in three steps: the raw data taking and pre-processing with EUDAQ \cite{eudaq}, the track and event reconstruction with EUTelescope \cite{eutelescope} and data analysis. The track and event reconstruction part includes grouping of the hits detected by the individual detectors to form the events, alignment of the telescope planes and fitting of the tracks. The data analysis is realised by a standalone Python script.
\par Even if the telescope system can reach an excellent spatial resolution, the achieved precision is not as good as expected from the pixel size of telescope detectors due to alignment issues. The effective resolution of the Mimosa detectors was degraded due to the alignment issues to approximately 10 $\mu$m. However, the achieved resolution is still sufficient for the purposes of this study.

\section{Analysis Results}
\label{analysis}
\subsection{Noise Determination}
\label{noisesec}
\par The baseline noise level is determined from the pedestal runs measured without the particle beam, so there is no induced current signal in the sensor. The pedestal data were measured both at $+30\:\degree\mathrm{C}$ and $-30\:\degree\mathrm{C}$ in the September 2019 test beam campaign. 

\par The noise level can be determined from the threshold scans with no beam present - the occupancy is measured as the threshold is scanned while a fixed signal amplitude is injected into the individual channels via internal circuitry. 
Two data sets are used for the determination of the noise.
The first data set did not use charge injection to the front-end amplifier, whereas the second data set used a fixed input charge injection of 0.2 fC.
The strip occupancy is defined as the fraction of events that had detected hits. As the threshold level is scanned from low to high values, initially all signals are recorded as hits. However, as the threshold approaches the signal level the occupancy starts to rapidly decrease. The sharp decrease of the occupancy is broadened by the noise. For a Gaussian distributed noise, the relation between occupancy and threshold is described as a complementary error function (often referred to as an ``S-curve''). The voltage level corresponding to the \mbox{50\%} occupancy point is called VT50. The gain is determined from the dependence of VT50 with injected signal amplitude. The output noise is determined from the Gaussian sigma of the S-curve. The input noise is defined as the output noise divided by the gain~\cite{spieler}~\cite{PeterThesis}.

\begin{figure*}
    \centering
     \begin{subfigure}[b]{0.48\textwidth}
    \includegraphics[width=\textwidth]{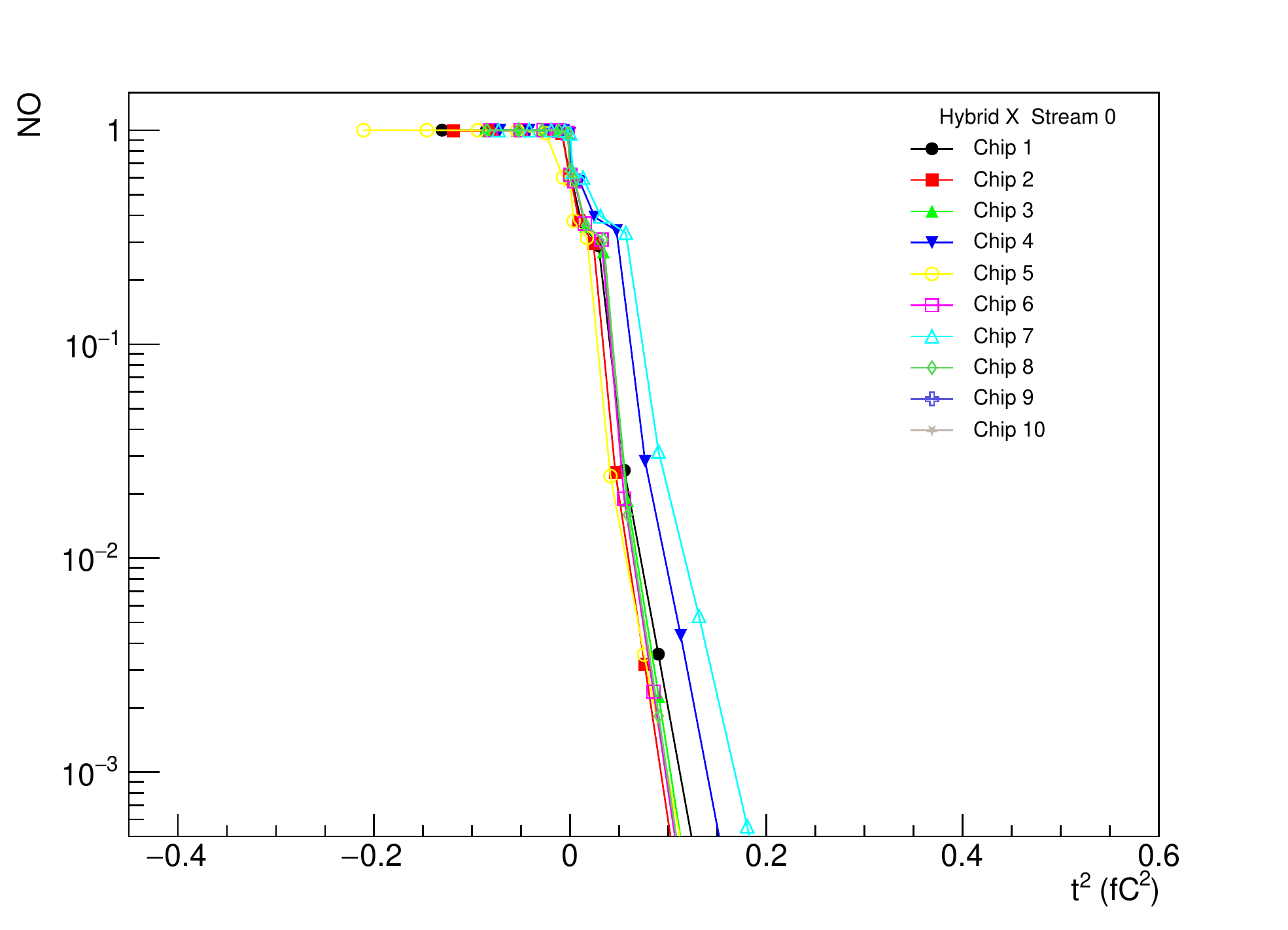}
    \caption{Hybrid X stream 0}
    \label{fig:plot1a}
    \end{subfigure}
     \hfill
    \begin{subfigure}[b]{0.48\textwidth}
    \includegraphics[width=\textwidth]{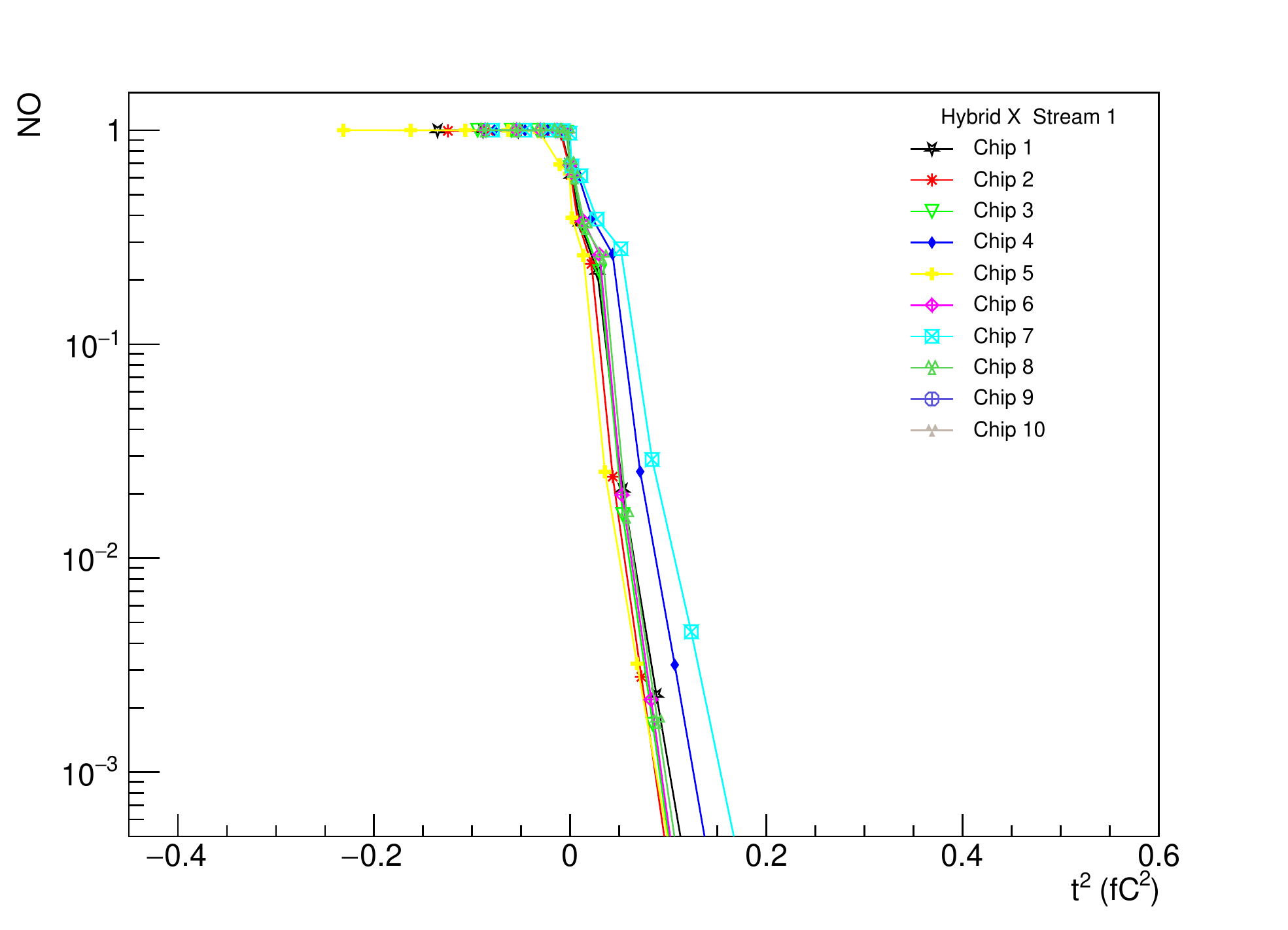}
    \caption{Hybrid X stream 1}
    \label{fig:plot1b}
    \end{subfigure}

     \begin{subfigure}[b]{0.48\textwidth}
    \includegraphics[width=\textwidth]{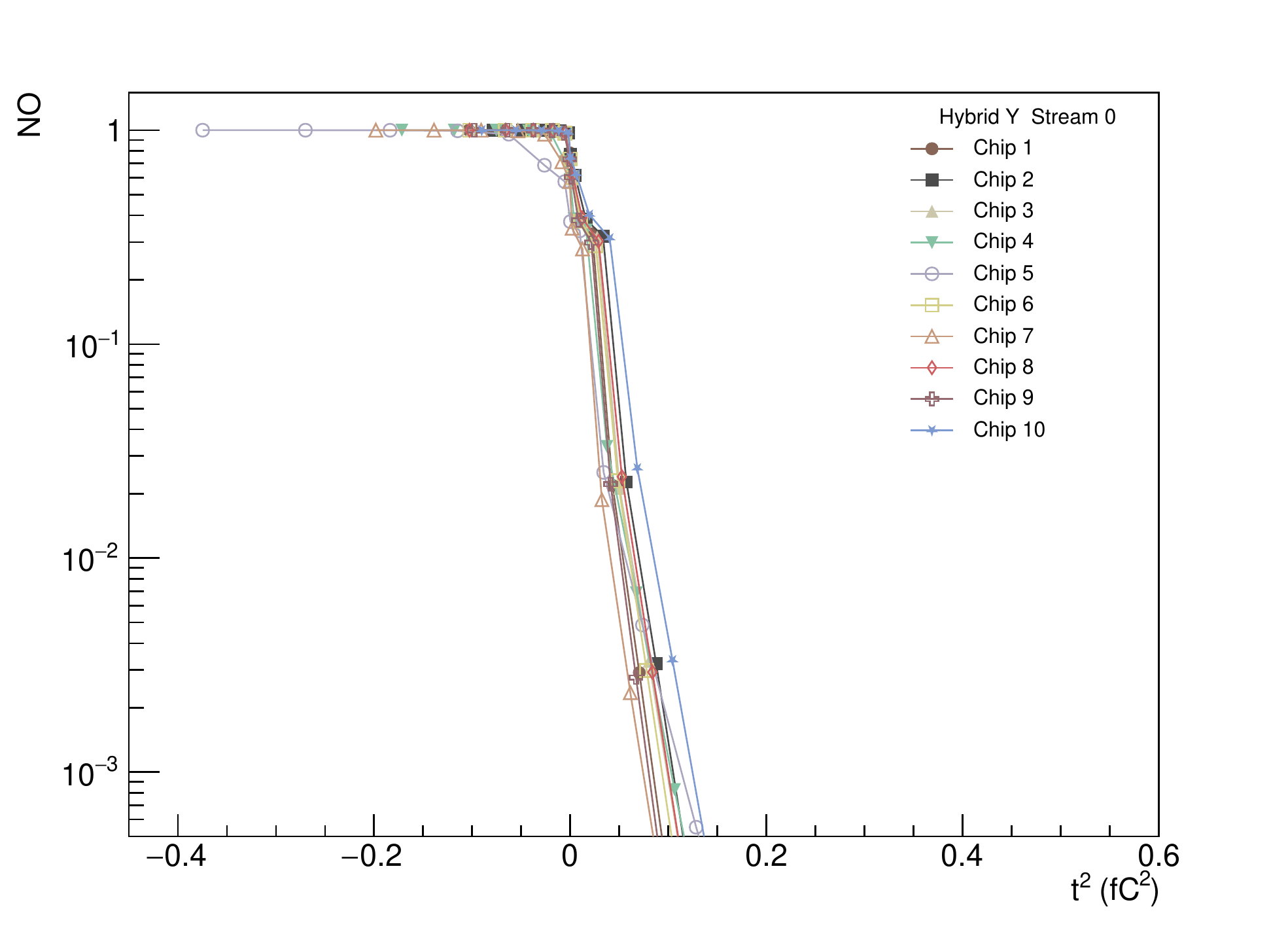}
    \caption{Hybrid Y stream 0}
    \label{fig:plot1c}
    \end{subfigure}
    \hfill
    \begin{subfigure}[b]{0.48\textwidth}
    \includegraphics[width=\textwidth]{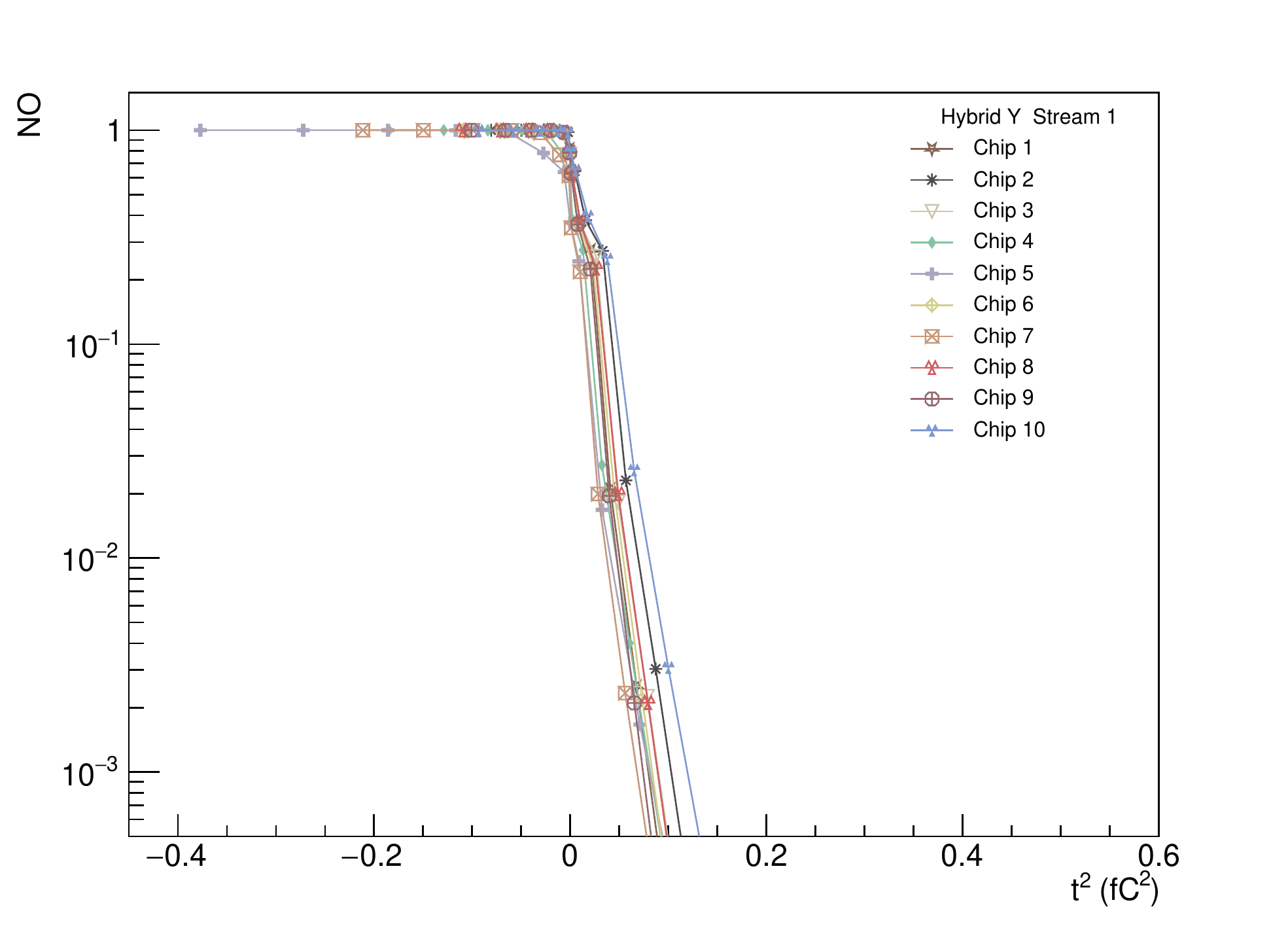}
    \caption{Hybrid Y stream 1}
    \label{fig:plot1d}
    \end{subfigure}
    \caption{Measured relation between Noise Occupancy ($NO$) and threshold$^2$ ($t^{2}$) for the cold measurements (at -30$\:\degree$C) without beam or injected charge. Hybrids X and Y are the two hybrids of the SS module. Streams 0 and 1 are the sensor sections to which the half-chips are wire-bonded.
     Data are shown for all ABCStar chips on hybrids X and Y.
    }
    \label{fig:plot1}
\end{figure*}

\begin{figure*}
    \centering
     \begin{subfigure}[b]{0.48\textwidth}
    \includegraphics[width=\textwidth]{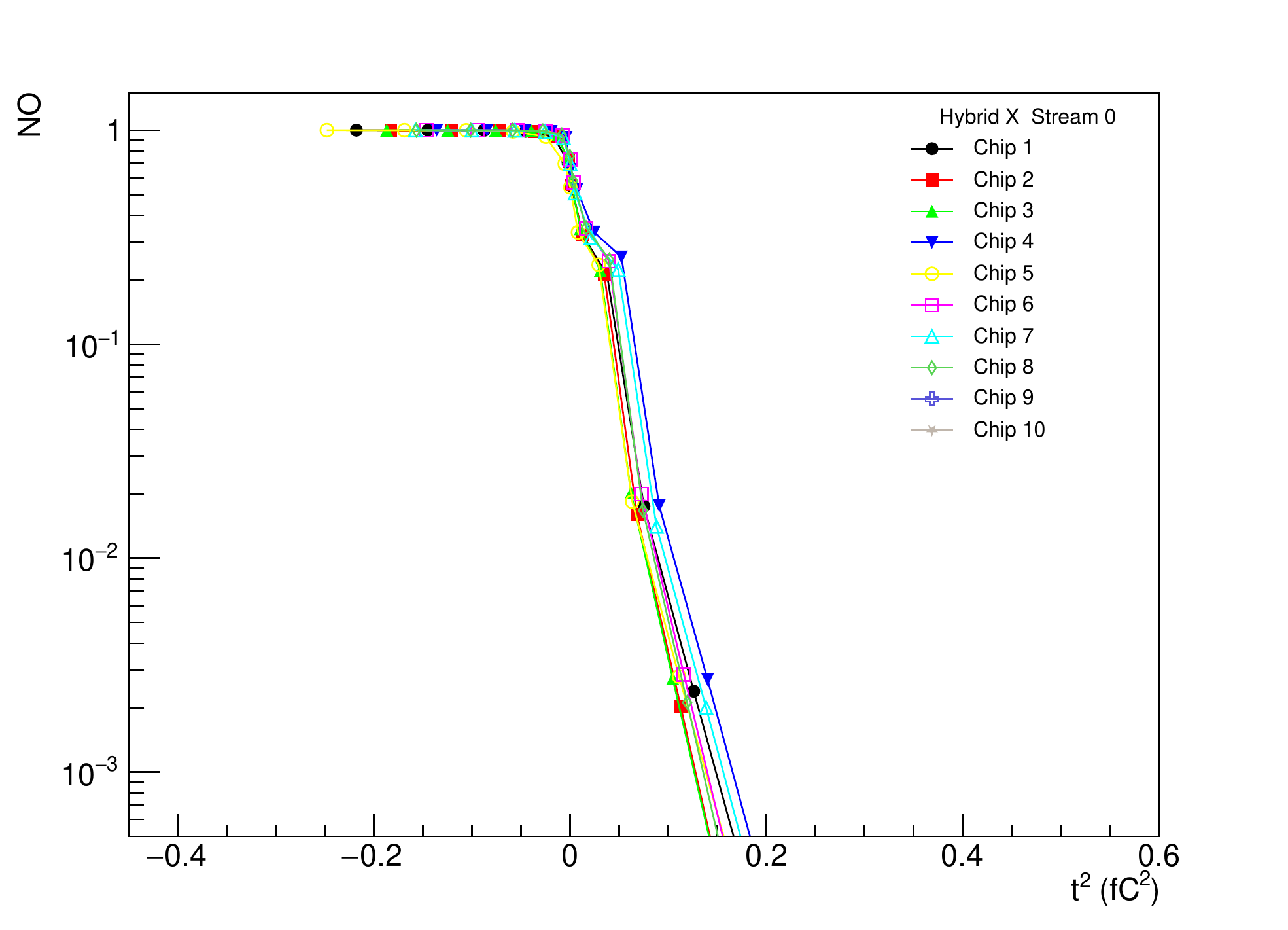}
    \caption{Hybrid X stream 0}
    \label{fig:plot18a}
    \end{subfigure}
     \hfill
    \begin{subfigure}[b]{0.48\textwidth}
    \includegraphics[width=\textwidth]{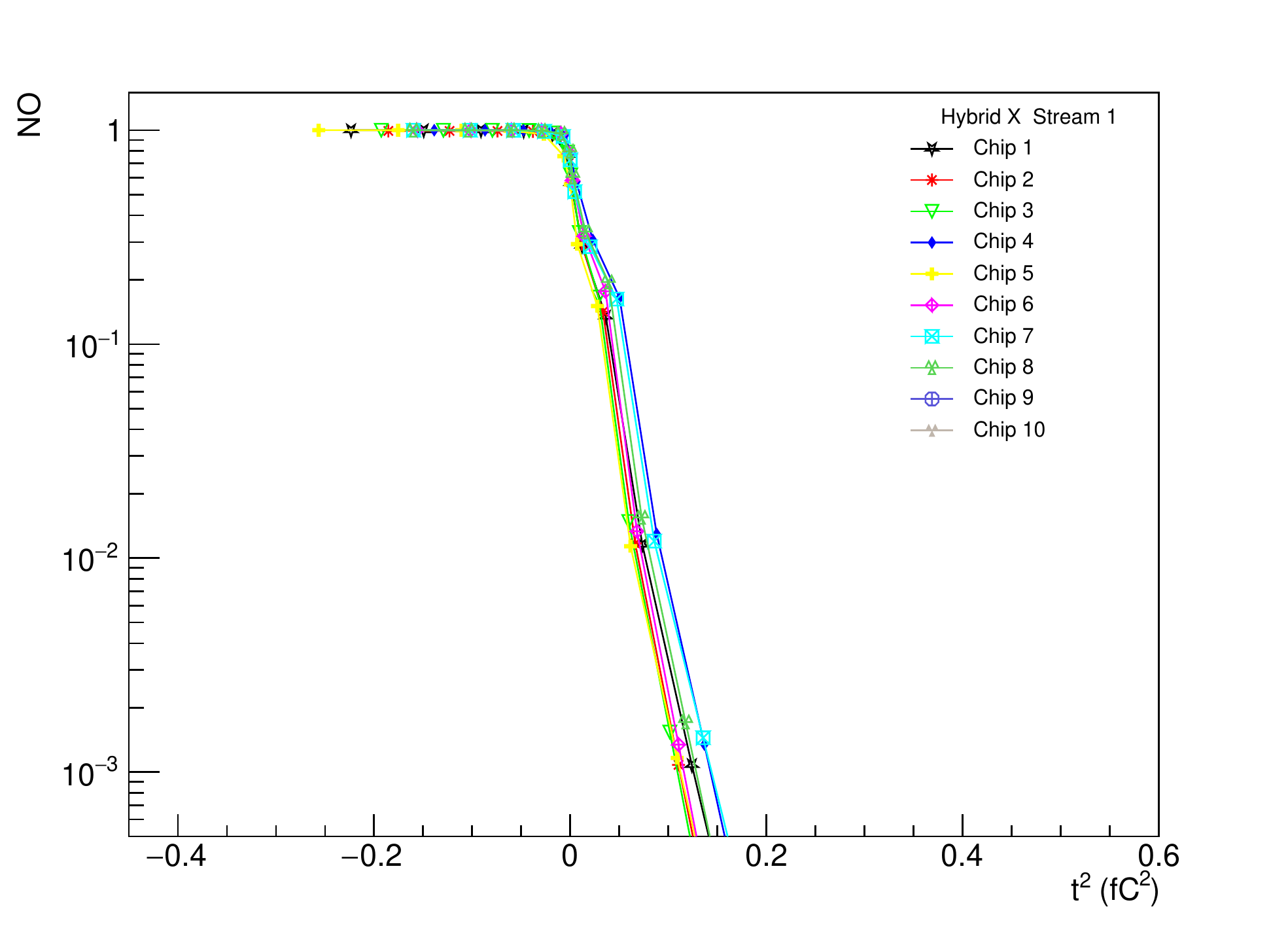}
    \caption{Hybrid X stream 1}
    \label{fig:plot18b}
    \end{subfigure}

     \begin{subfigure}[b]{0.48\textwidth}
    \includegraphics[width=\textwidth]{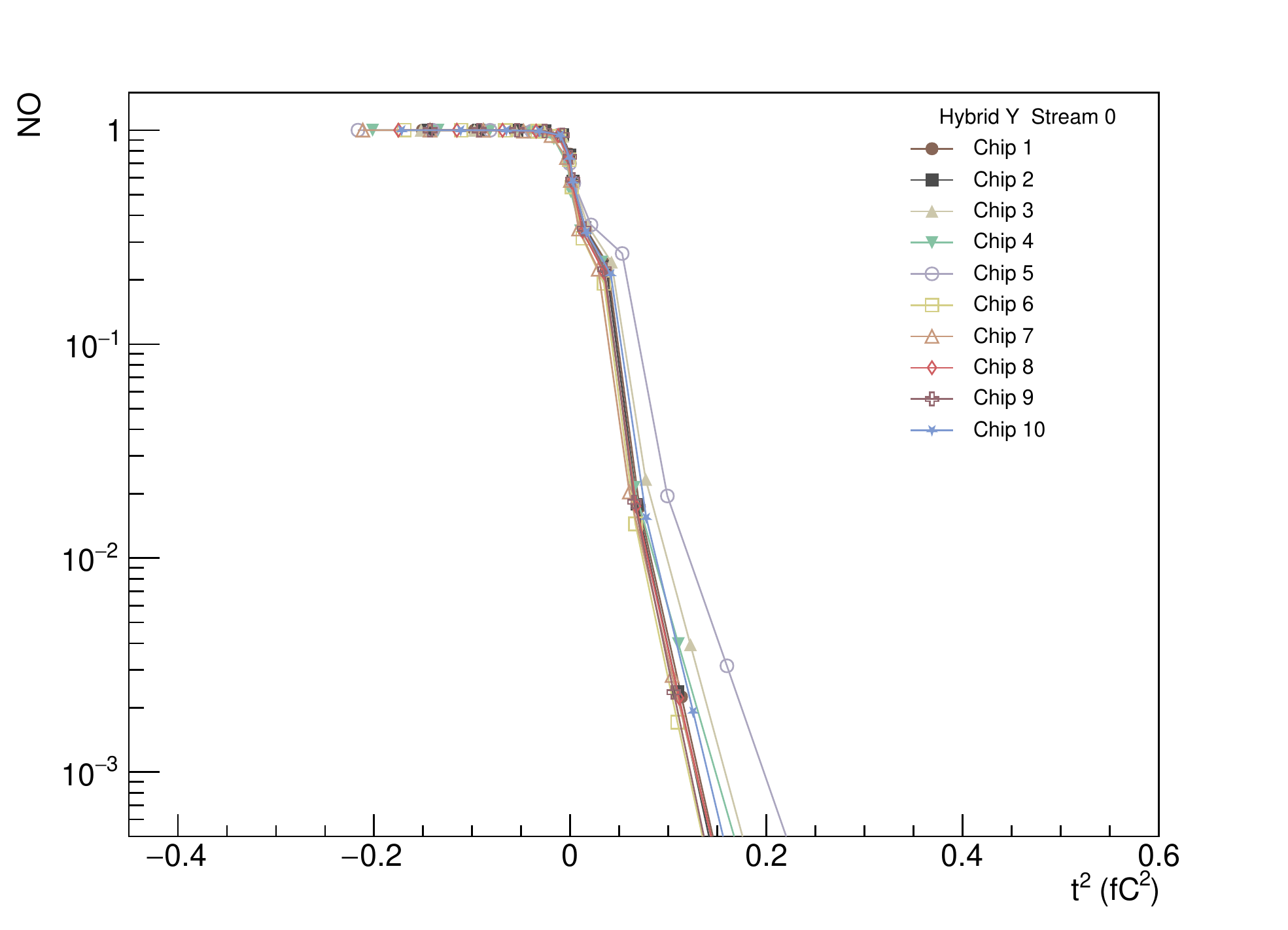}
    \caption{Hybrid Y stream 0}
    \label{fig:plot18c}
    \end{subfigure}
    \hfill
    \begin{subfigure}[b]{0.48\textwidth}
    \includegraphics[width=\textwidth]{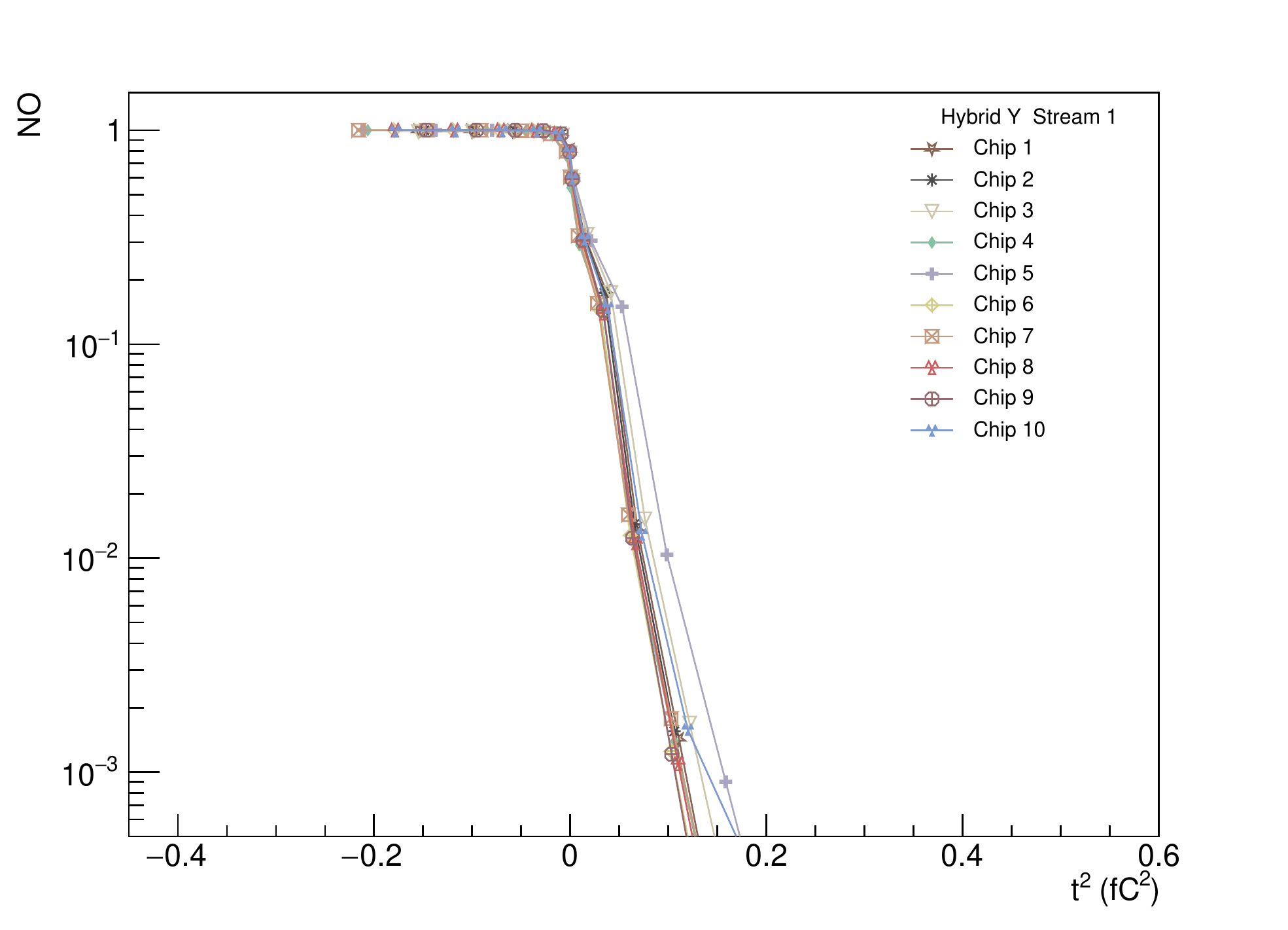}
    \caption{Hybrid Y stream 1}
    \label{fig:plot18d}
    \end{subfigure}
    \caption{Measured relation between Noise Occupancy ($NO$) and threshold$^2$ ($t^{2}$) for the warm measurements (at +30$\:\degree$C) without beam or injected charge. Hybrids X and Y are the two hybrids of the SS module. Streams 0 and 1 are the sensor sections to which the half-chips are wire-bonded.
     Data are shown for all ABCStar chips on hybrids X and Y.
    }
    \label{fig:plot18}
\end{figure*}

For each strip of the module, the dependence of the occupancy on the threshold is measured and fitted with a complementary error function. With no input charge, the pedestal hits are mainly caused by the electronic noise. Half of an ABCStar chip is wire bonded to the same row of strips (see Figure~\ref{fig:plot13}).  The performance of the half chip is characterized by the S-curve averaged over 128 corresponding channels. The noise occupancy ($NO$) and the signed square of the threshold ($t^2$) are obtained. The logarithmic relation between $NO$ and $t^2$ (unit in fC$^{2}$) is illustrated in Figures~\ref{fig:plot1} and~\ref{fig:plot18} for the cold and warm measurements, respectively. The plateau in Figure \ref{fig:plot1} extends to negative charge, which is an artefact of the calibration~\cite{ref2}. As the S-curves are proportional to a complementary error function, the relation between the noise occupancy in log scale and the signed square of the threshold is expected to be linear in the falling part, which can be seen from Figure~\ref{fig:plot1} and Figure~\ref{fig:plot18}.

The electrical noise level of the readout channel is characterized by the Gaussian distribution associated with the S-curve measured for this channel. The procedure used determines the output noise of the amplifier in mV and this is converted into the input noise in fC by allowing for the gain. 

\par
\begin{figure*}
    \centering
    \begin{subfigure}[b]{\textwidth}
    \centering
        \includegraphics[width=0.8\textwidth]{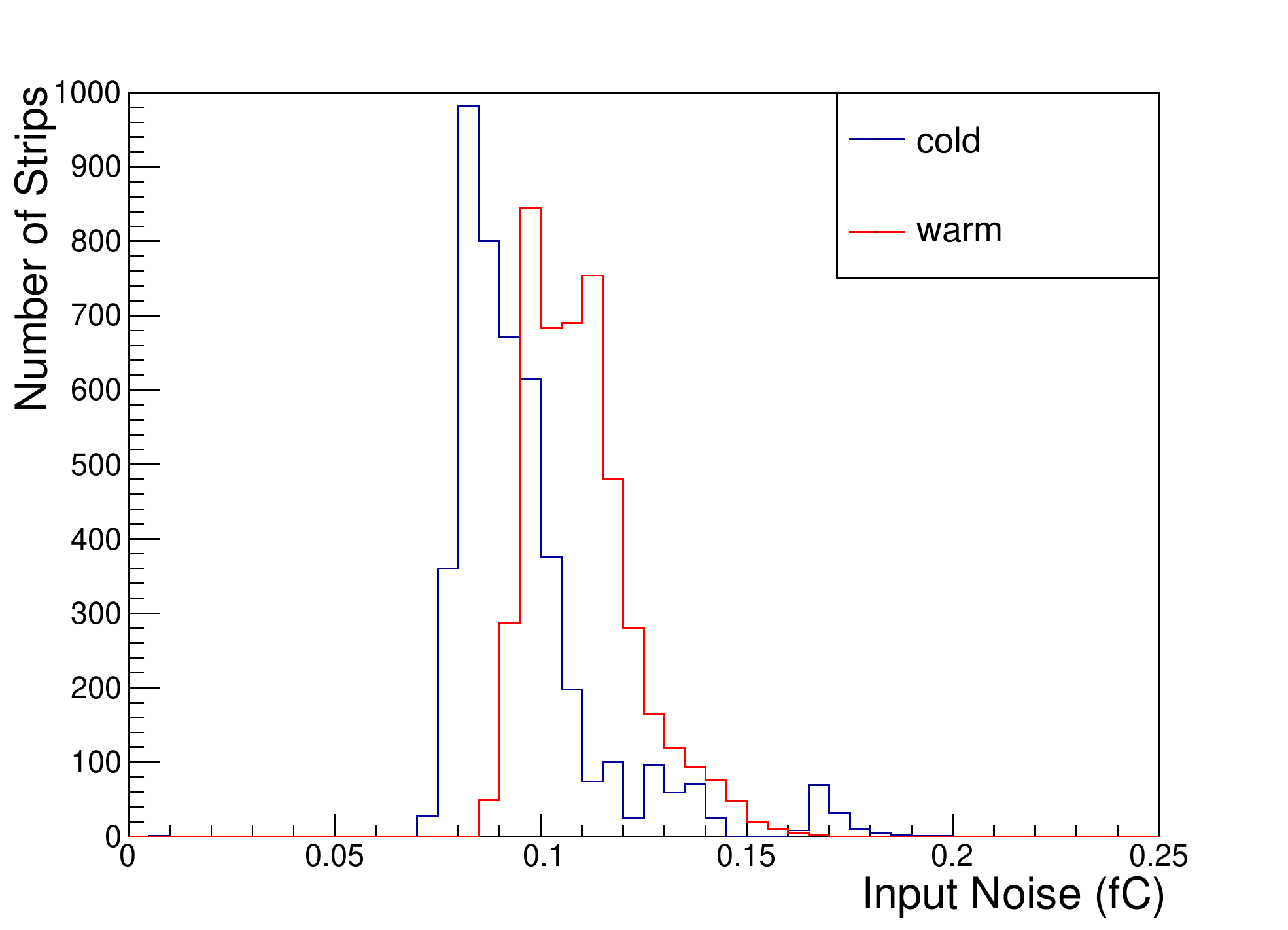}
        \caption{Injected calibration charge 0 fC}
    \label{fig:plot2a}
    \end{subfigure}
    \begin{subfigure}[b]{\textwidth}
    \centering
        \includegraphics[width=0.8\textwidth]{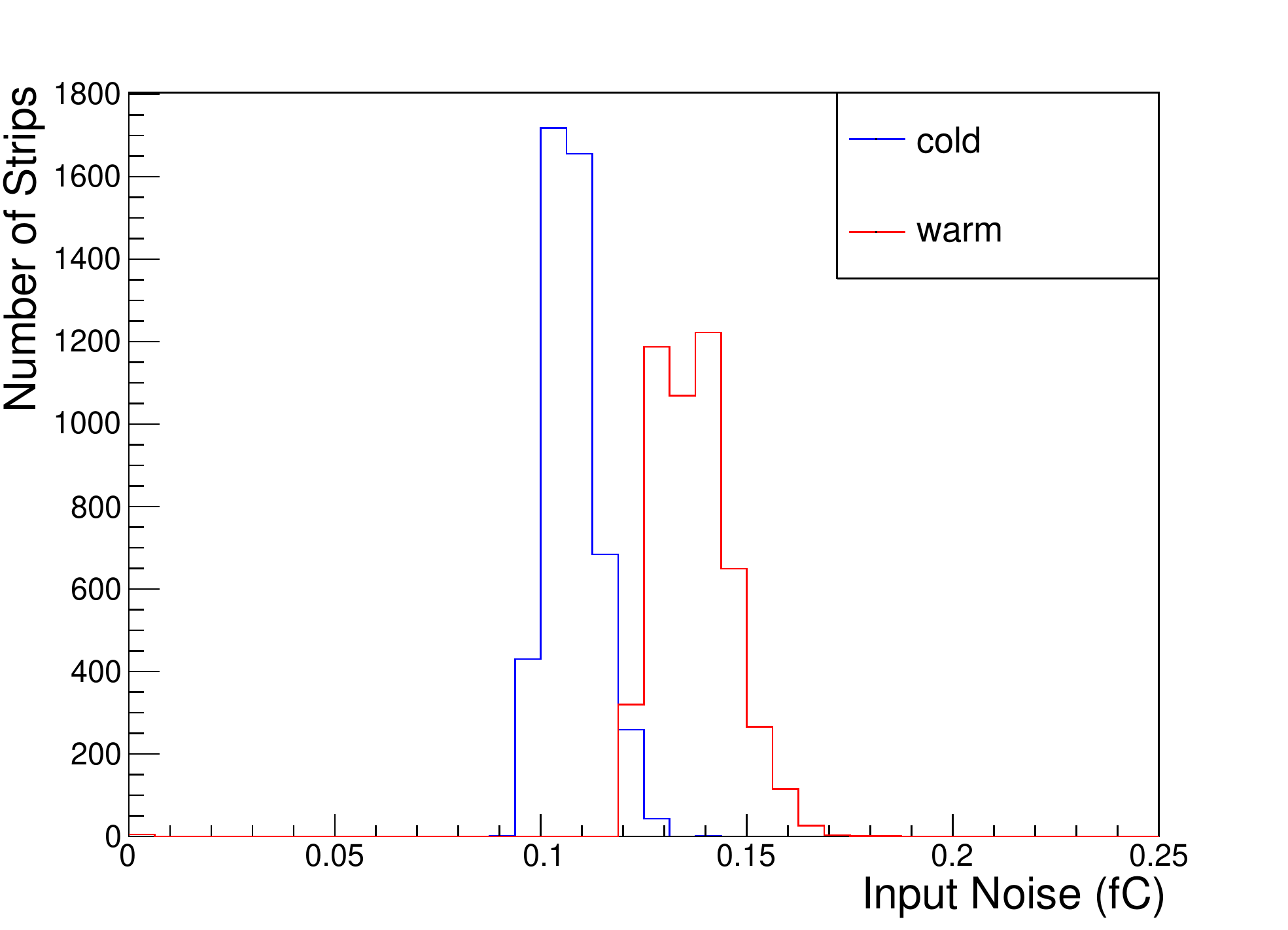}
        \caption{Injected calibration charge 0.2 fC}
        \label{fig:plot2b}
    \end{subfigure}
    \label{plot2}
    \caption{The input noise measured for individual channels of the ITk SS module at $-30\:\degree\mathrm{C}$ (blue solid line) and $+30\:\degree\mathrm{C}$ (red solid line). a) Input calibration charge 0 fC. The mean value of the cold distribution is 0.0958 fC with a standard deviation of 0.0185 fC. The mean value of the warm distribution is 0.1099 fC with a standard deviation of 0.0128 fC. b) Input calibration charge 0.2 fC. The mean value of the cold distribution is 0.0859 fC with a standard deviation of 0.0058 fC.  The mean value of the warm distribution is 0.1097 fC with a standard deviation of 0.0080 fC.}
\end{figure*}

 For 0 fC injection charge, Figure~\ref{fig:plot2a} shows the measured input noise (fC) of each individual channel, for operating temperatures of respectively $-30\:\degree\mathrm{C}$ and $+30\:\degree\mathrm{C}$. The noise was measured for channels read out by the 18 ABCStar ASICs that were used. There were 3-4 dead or noisy channels removed from the analysis out of 4608 channels. No channels are excluded from the 123 strips used for the analysis of the beam data. The distributions are fitted by Gaussian functions and the obtained mean values correspond to the electrical noise level. Three different estimates of mean noise were used for each distribution in Figure~\ref{fig:plot2a}; the mean of the distribution, the mean of the Gaussian fit of the distribution and the truncated mean.  The cold peak in Figure~\ref{fig:plot2a} at 0.17 fC is an artifact of the calibration, thus excluded from the Gaussian fit and the truncated mean calculations.
The mean value of the cold distribution is 0.0957 fC with a standard deviation of 0.0185 fC. The uncertainty on the mean is 0.0003 fC. The mean value of the warm distribution is 0.1099 fC with a standard deviation of 0.0128 fC. The uncertainty on the mean is 0.0002 fC.

\par The second data set used to measure the electrical noise level is acquired by injecting the front-end with a delta pulse of 0.2 fC from the internal calibration circuit. The distribution of the input noise obtained is shown in Figure~\ref{fig:plot2b}.  When the charge injection is at 0.2 fC, the mean value of the cold distribution is 0.0859 fC with a standard deviation of 0.058 fC. The uncertainty on the mean is 0.0001 fC. The mean value of the warm distribution is 0.1097 fC with a standard deviation of 0.0080 fC. The uncertainty on the mean is 0.0001 fC. The same procedures are applied to the second data set to obtain three estimates.  The six estimators obtained from each method and data set are summarised in Tab.~\ref{tab:tab3}.

\begin{table}[]
\resizebox{\textwidth}{!}{%
\begin{tabular}{|l|ccccc|ccccc|}
\hline
                  & \multicolumn{5}{c|}{0 fC}                                                                                                                                                                                                                                                                                                                    & \multicolumn{5}{c|}{0.2 fC}                                                                                                                                                                                                                                                                                                                  \\ \hline
                  & \multicolumn{1}{l|}{\begin{tabular}[c]{@{}l@{}}cold \\ (ENC)\end{tabular}} & \multicolumn{1}{l|}{\begin{tabular}[c]{@{}l@{}}cold \\ (fC)\end{tabular}} & \multicolumn{1}{l|}{\begin{tabular}[c]{@{}l@{}}warm \\ (ENC)\end{tabular}} & \multicolumn{1}{l|}{\begin{tabular}[c]{@{}l@{}}warm \\ (fC)\end{tabular}} & \multicolumn{1}{l|}{ratio} & \multicolumn{1}{l|}{\begin{tabular}[c]{@{}l@{}}cold \\ (ENC)\end{tabular}} & \multicolumn{1}{l|}{\begin{tabular}[c]{@{}l@{}}cold \\ (fC)\end{tabular}} & \multicolumn{1}{l|}{\begin{tabular}[c]{@{}l@{}}warm \\ (ENC)\end{tabular}} & \multicolumn{1}{l|}{\begin{tabular}[c]{@{}l@{}}warm \\ (fC)\end{tabular}} & \multicolumn{1}{l|}{ratio} \\ \hline
Raw mean          & \multicolumn{1}{c|}{598}                                                   & \multicolumn{1}{c|}{0.0958}                                               & \multicolumn{1}{c|}{686}                                                   & \multicolumn{1}{c|}{0.1099}                                               & 1.147                      & \multicolumn{1}{c|}{536}                                                   & \multicolumn{1}{c|}{0.0859}                                               & \multicolumn{1}{c|}{685}                                                   & \multicolumn{1}{c|}{0.1097}                                               & 1.280                      \\ \hline
Gaussian fit mean & \multicolumn{1}{c|}{568}                                                   & \multicolumn{1}{c|}{0.0910}                                               & \multicolumn{1}{c|}{674}                                                   & \multicolumn{1}{c|}{0.1080}                                               & 1.187                      & \multicolumn{1}{c|}{536}                                                   & \multicolumn{1}{c|}{0.0859}                                               & \multicolumn{1}{c|}{685}                                                   & \multicolumn{1}{c|}{0.1097}                                               & 1.278                      \\ \hline
Truncated mean    & \multicolumn{1}{c|}{564}                                                   & \multicolumn{1}{c|}{0.0904}                                               & \multicolumn{1}{c|}{669}                                                   & \multicolumn{1}{c|}{0.1072}                                               & 1.186                      & \multicolumn{1}{c|}{536}                                                   & \multicolumn{1}{c|}{0.0859}                                               & \multicolumn{1}{c|}{678}                                                   & \multicolumn{1}{c|}{0.1086}                                               & 1.266                      \\ \hline
\end{tabular}%
}
\caption{Three estimates of the mean input noise (fC and ENC) measured for cold
($-30\:\degree\mathrm{C}$) and warm ($+30\:\degree\mathrm{C}$) operating temperatures, and respectively 0 fC
and 0.2 fC injected charge. The unit ENC is the equivalent noise charge, which is the noise charge measured in electrons.}
\label{tab:tab3}
\end{table}

\par Since the Gaussian fit gives a better estimate of the peak, the Gaussian fit means at 0 and 0.2 fC are taken to give the final noise.  The difference between the two Gaussian means is given as a conservative estimate of the uncertainties. The average noise of the two estimates at $-30\:\degree\mathrm{C}$ is $552\pm32\;e^{-}$, while the average noise at $+30\:\degree\mathrm{C}$ is $680\pm11\;e^{-}$. Tab. \ref{tab:tab3} shows the results of the six estimates of the ratio between the noise measured warm and cold. Taking the average of the two Gaussian measurements gives an increase of the noise between the cold and warm measurements of $23.3\%\pm9.1\%$. The noise differences between 0 and 0.2 fC charge injections are systematic and not fully understood. 

The temperature affects the series and the parallel noise of the pre-amplifier. For the series noise the transconductance ($g_m$) decreases with increasing temperature (T)~\cite{ref1}. The series noise scales as $\sqrt{T}/g_m$)~\cite{Kaplon}. The estimated difference in this noise source between cold and warm temperatures ($-30\degree\mathrm{C}$ and $+30\degree\mathrm{C}$)  is about 22\%. 
The parallel noise scales as $\sqrt{T/R_f}$, where $R_f$ is the resistance of the feedback resistor~\cite{Kaplon}. This results in an increase in the series noise between cold and warm temperatures of about 12\%. Using the measured noise contribution as a function of the input capacitance~\cite{ref2}, the overall increase in the noise is estimated to be 17\%.
This result is thus qualitatively compatible with the calculation given the experimental systematic uncertainties and the uncertainties on the temperature measurements.

\par By fitting the distribution of the signal values - thresholds for which the noise occupancy decreases to the half of its original value - obtained from the S-curves of individual channels measured without the beam by the Gaussian function, we obtain the offset from the 0 fC, which needs to be subtracted from the measurements with the beam.

\subsection{Characterization of the module with particle beam}
The charge measured by the DUT is discussed and compared with the simulation in Sec.~\ref{charge}. The signal-to-noise ratio (SNR) for the unirradiated DUT is obtained. The cluster sizes are discussed in Sec.~\ref{cluster}, while the pulse shape reconstruction is studied in Sec.~\ref{pulseshape}. 
The number of entries in the measured charge studies is determined by the number of strips hit by the beam (see Figure \ref{fig:plot14}).

\subsubsection{Measured Charge}
\label{charge}
The ITk SS module was tested at operational temperatures of $-30\:\degree\mathrm{C}$ and $+30\:\degree\mathrm{C}$ by the 5.4 GeV electron beam described in Sec. \ref{desybeam} produced by the DESY-II synchrotron. The detection efficiency, defined as 
\begin{equation}
    \mathrm{efficiency} = \frac{track_\mathrm{DUT+FEI4}}{track_\mathrm{FEI4}}
    \label{Eq:eq1}
\end{equation}
\noindent is an important figure of merit for the module performance. In practice, the DUT is considered to be efficient in a given event if the hit is detected by the DUT within 3 strips from the particle track reconstructed from the Mimosa26 data.

\begin{figure}
    \centering
    \includegraphics[width=0.8\textwidth]{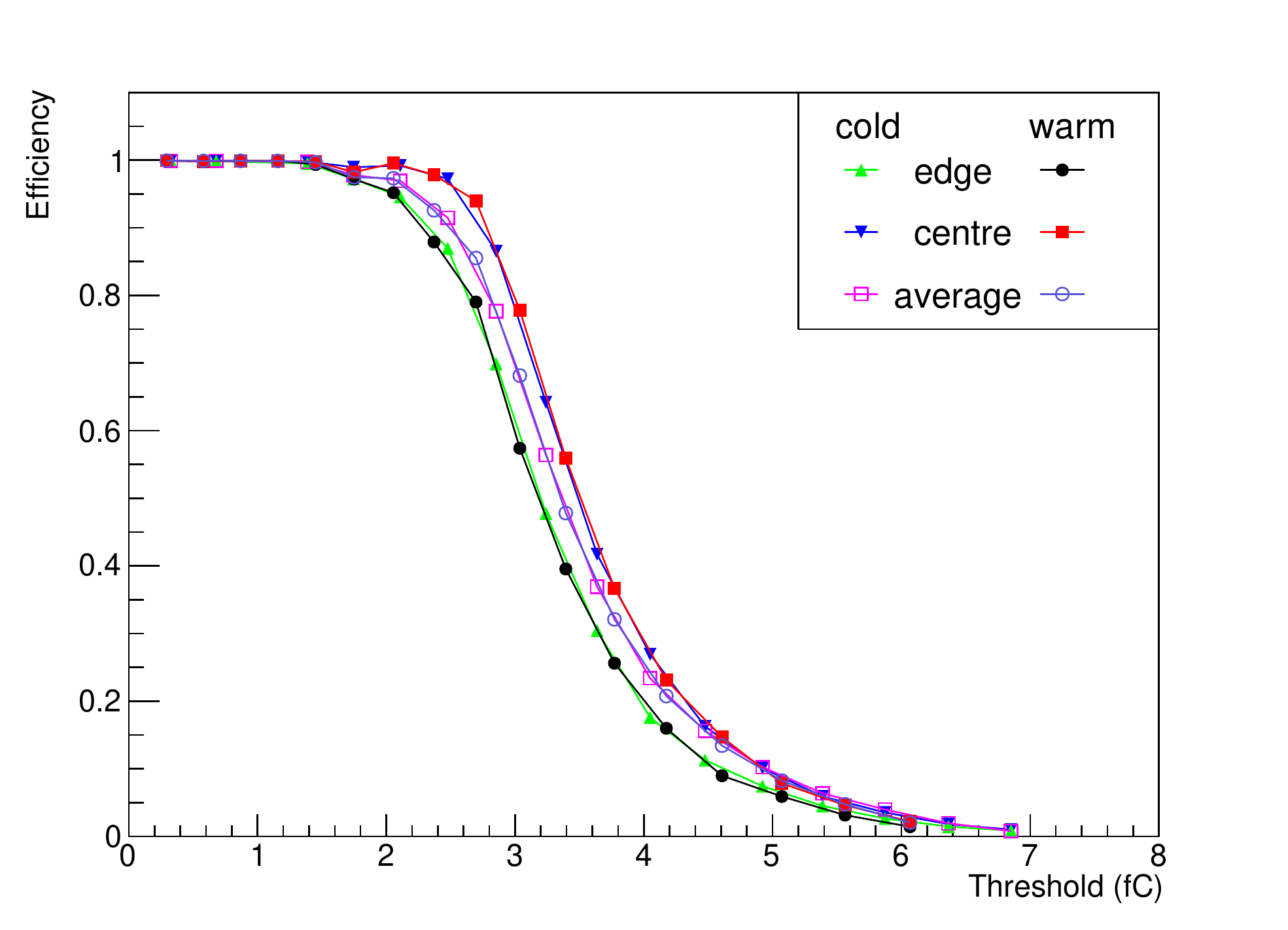}
    \caption{Detection efficiency of the ITk SS module measured cold ( $-30\:\degree\mathrm{C}$) and warm ($+30\:\degree\mathrm{C}$).}
    \label{fig:plot3}
\end{figure}

The position where the beam particle hits the strip can cause a difference in efficiency. If the particle passes the centre of the sensor strip, it is more likely that most of the induced current signal is generated on this strip and a hit is recorded. However, if the particle hits the edge of the strip, the induced current signal is more likely to be shared between adjacent strips. Each strip then detects a smaller signal and thus being less likely to record a hit above the threshold. This effect is displayed in Figure~\ref{fig:plot3}, in which ``centre'' is defined as the 20\% region of the strip around the centre of the strip while ``edge'' is defined as the 20\% area closest  to the edge, and ``average'' covers the full width of the strip. The performance in the three regions are averaged over all the strips in the beam region and the results are shown in Figure~\ref{fig:plot3}. It can be seen that for the same threshold, the efficiency is lower when the track passes through the strip edge and higher when the track passes the centre.
However, in HL-LHC operation the threshold will be around 1 fC and therefore the loss of efficiency from the "edge" region will be negligible (see Figure~\ref{fig:plot3}). 

\par
\begin{figure}
    \centering
    \includegraphics[width=0.8\textwidth]{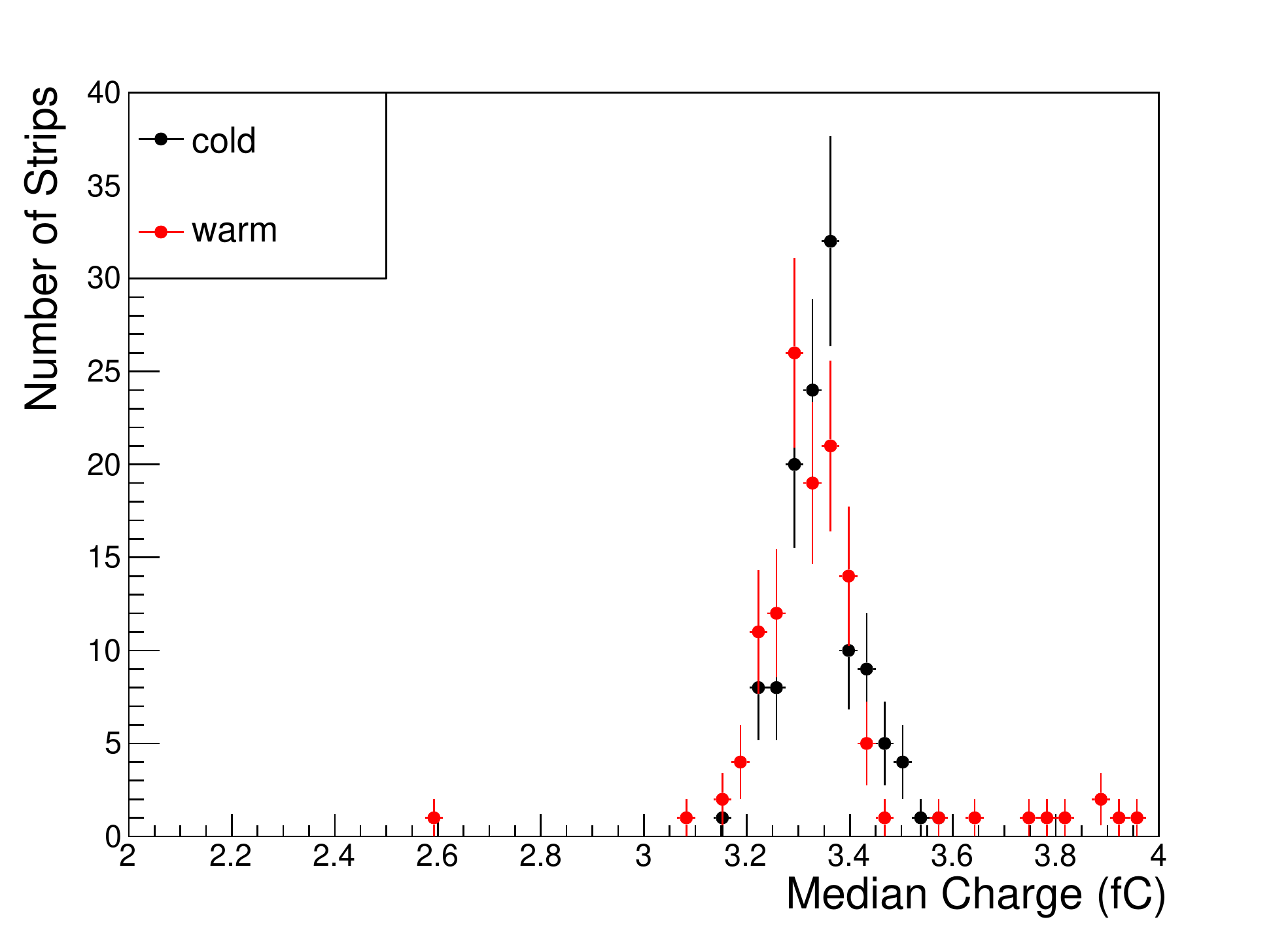}
    \caption{Distribution of the values of median charge extracted from the S-curves measured for each individual channel of the ITk SS module both at $-30\:\degree\mathrm{C}$ (black) and $+30\:\degree\mathrm{C}$ (red) after subtracting the offset from the 0 fC (as discussed in Sec.~\ref{noisesec}). The mean value for the cold distribution is 3.346 fC with a standard deviation of 0.072 fC. The mean value for the warm distribution is 3.344 fC with a standard deviation of 0.160 fC. The statistical uncertainty on the mean is respectively 0.007 fC and 0.014 fC for the cold and warm distributions.}
    \label{fig:plot7}
\end{figure}

\begin{figure}
    \centering
    \includegraphics[width=0.8\textwidth]{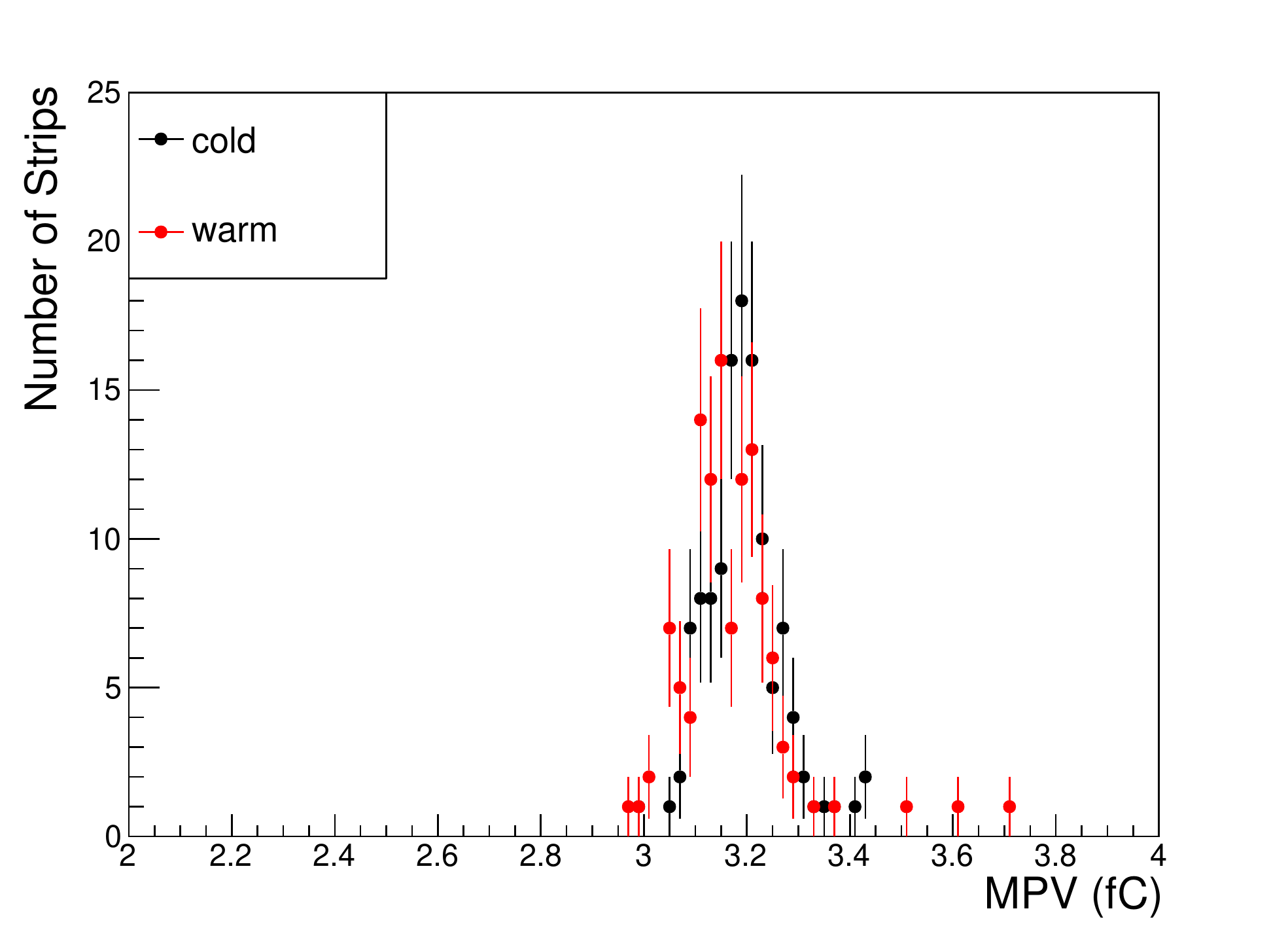}
    \caption{Distribution of the most probable values (MPV), corresponding to the measured charge, calculated for the S-curves measured for each individual channel of the ITk SS module both at $-30\:\degree\mathrm{C}$ (black) and $+30\:\degree\mathrm{C}$ (red) after subtracting the offset from the 0 fC (as discussed in Sec.~\ref{noisesec}). The mean value for the cold distribution is 3.19 fC with a standard deviation of 0.08 fC. The mean value for the warm distribution is 3.18 fC with a standard deviation of 0.16 fC. The statistical uncertainty on the mean is respectively 0.007 fC and 0.014 fC for the cold and warm distributions. }
    \label{fig:plot4}
\end{figure}

 The shape of the signal from the charged particle passing the sensor follows the Landau distribution, while the noise distribution is Gaussian. The S-curve measured with the beam for each individual channel is fitted by the empirical skewed complementary error function defined by the equation~\cite{ref3}, 
\begin{equation}
    f(x) = E_{0} \cdot \mathrm{erfc}\left(\alpha \cdot \left(1 + 0.6 \cdot \frac{ e^{-A \alpha }-e^{ A\alpha }}{e^{ -A\alpha }+e^{ A\alpha }}\right) \right),\;\; \alpha = \frac{x-x_{0}}{\sqrt{2}\sigma} 
    \label{Eq:eq2}
\end{equation}

\noindent where $E_{0}$ means maximum efficiency, which is considered to be 1 in our case, and $x_{0}$ is the median charge corresponding to the threshold for which the detection efficiency decreases to 50\%. The $\sigma$ is the fit parameter correlated with the width of the distribution and $A$ is the skewed parameter. The distribution of the values of median charge measured for all 123 readout channels after subtracting the offset from the 0 fC (as discussed in Sec.~\ref{noisesec}) is shown in Figure~\ref{fig:plot7}. The warm outliers come from the failed or inaccurate fits.  The signal charge measurement corresponding to the particular strip is represented by the most probable value (MPV) obtained by a numerical differentiation of the function fitting the S-curve measured for this strip. As shown in Figure~\ref{fig:plot4}, the mean value of the distribution of MPVs of all the strips after subtracting the offset from the 0 fC measured at $-30\:\degree\mathrm{C}$ is $3.19\;\mathrm{fC}$, while it is $3.18\;\mathrm{fC}$ for the measurement at $+30\:\degree\mathrm{C}$. Both systematic and statistical uncertainties are considered.  Sources of systematic uncertainties include the uncertainty introduced by the fit function, the definition of the strip centre, the cross-talk measurements~\cite{ref3}, as well as the uncertainty introduced by calibration~\cite{ref2}. 
The uncertainty from the definition of the ``strip centre'' was estimated by changing the definition of the strip centre from including 20 \% to including 40 \% of the strip.  
The total combined MPV uncertainty is about \mbox{0.17 fC} for both cold and warm measurements. The contributions from different sources of uncertainty are given in Tab.~\ref{tab:tab4}. The MPVs extracted from the cold and warm measurements are quite close, agreeing with the signal being independent of the module temperature.

\begin{table}[]
\centering
\begin{tabular}{|c|c|c|c|c|c|}
\hline
     & Fit Function (fC) & \begin{tabular}[c]{@{}c@{}}Centreing\\ (20\% vs 40\%) (fC)\end{tabular} & Cross Talk (fC) & Calibration (fC) & Statistical (fC) \\ \hline
Cold & 0.054        & 0.014                                                              & 7.24 $\times$ $10^{-4}$      & 0.160       & 0.007       \\ \hline
Warm & 0.038        & 0.02                                                               & 7.24 $\times$ $10^{-4}$       & 0.159       & 0.014       \\ \hline
\end{tabular}
\caption{Contributions to the measurement uncertainty of the signal MPV (fC), for cold ($-30\:\degree\mathrm{C}$) and warm ($+30\:\degree\mathrm{C}$) operation.}
\label{tab:tab4}
\end{table}

\par
\begin{figure}
    \centering
    \includegraphics[width=\textwidth]{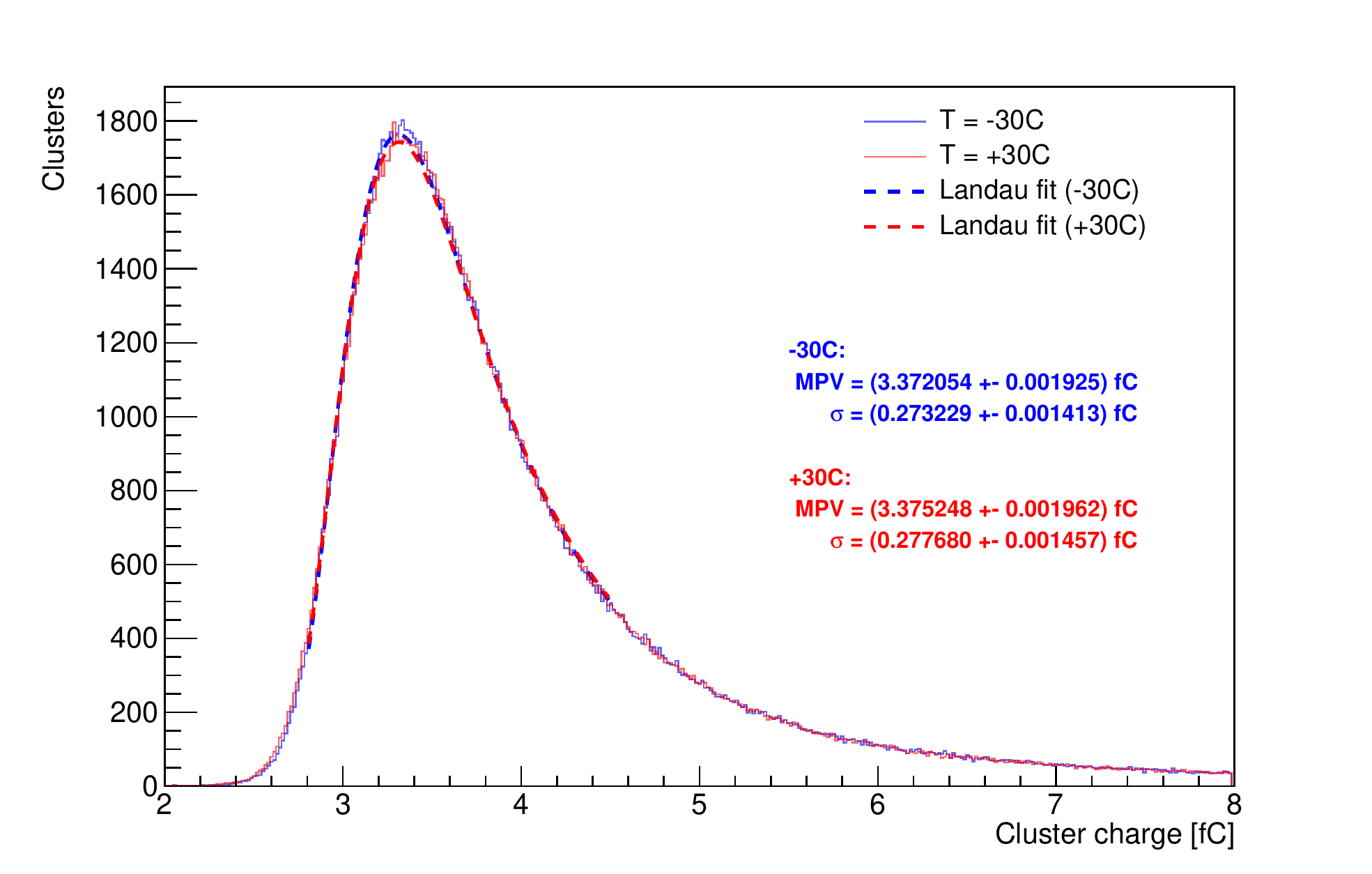}
    \caption{Distributions of cluster charge obtained from an Allpix-Squared ~\cite{allpix} for two temperature settings ($-30\:\degree\mathrm{C}$ and at $+30\:\degree\mathrm{C}$) using 5.4 GeV electrons.}
    \label{fig:plot19}
\end{figure}

 Several factors can reduce the measured charge signal. The first factor relates to the fact that data accounts for all events regardless of the exact position of the particle passing through the silicon sensor. In an ideal case of the particles passing through the strip centre the measured charge signal would be higher, see Figure~\ref{fig:plot3}. This effect was included by defining a correction factor based on the ratio of MPV values determined from strip centre and average. Another source of the signal loss is cross talk~\cite{ref3}, which accounts for approximately 4~\% of signal loss. A Monte-Carlo simulation in the Allpix-Squared~\cite{allpix} framework, utilizing GEANT4~\cite{GEANT4}, is used to simulate the detector response to electrons with the energy of 5.4 GeV.
The simulation has taken the cross talk factor into consideration, with the result shown in Figure~\ref{fig:plot19}. The simulated MPVs are 3.37 fC at $-30\:\degree\mathrm{C}$ and 3.38 fC at $+30\:\degree\mathrm{C}$. 
After correcting the data for the centre  and edge effects, the measured MPVs are 3.37 fC at $-30\:\degree\mathrm{C}$ and 3.34 fC at $+30\:\degree\mathrm{C}$, see Tab.~\ref{Tab:tab2}. 
The simulation agrees with the data within uncertainties.  The ratio between MPVs after correction at $-30\:\degree\mathrm{C}$ and at $+30\:\degree\mathrm{C}$ is 1.009 $\pm$ 0.020. The dominant systematic uncertainty is from the calibration, which is highly correlated between cold and warm instances. Therefore the uncertainty on the MPV ratio is calculated using the other uncertainties in Tab.~\ref{tab:tab4}. It is compatible with unity as expected.


\begin{table}[h!]
\begin{center}
\begin{tabular}{ |p{0.05\linewidth}|p{0.13\linewidth}|p{0.12\linewidth}|p{0.1\linewidth}|
p{0.15\linewidth}|p{0.15\linewidth}|p{0.12\linewidth}|} 
\hline
 & Simulation charge (fC) & Uncorrected charge (fC) & Corrected charge (fC) & Overall Uncertainty (fC) & Calibration Uncertainty (fC) & Other Uncertainties (fC) \\
\hline
Cold & 3.37 & 3.19 & 3.37 & 0.17 & 0.16 & 0.06\\ 
Warm & 3.38 & 3.18 & 3.34 & 0.17 & 0.16 & 0.06\\ 
\hline
\end{tabular}
\caption{Comparison of the mean of the most probable (MPV) charge measured in the ITK Short Strip module using the 5.4 GeV electron beam - both uncorrected and corrected charge values are listed - with the output of the GEANT4 simulation. All values are listed in units of fC.}
\label{Tab:tab2}
\end{center}
\end{table}

\par
\begin{figure}
    \centering
    \includegraphics[width=\textwidth]{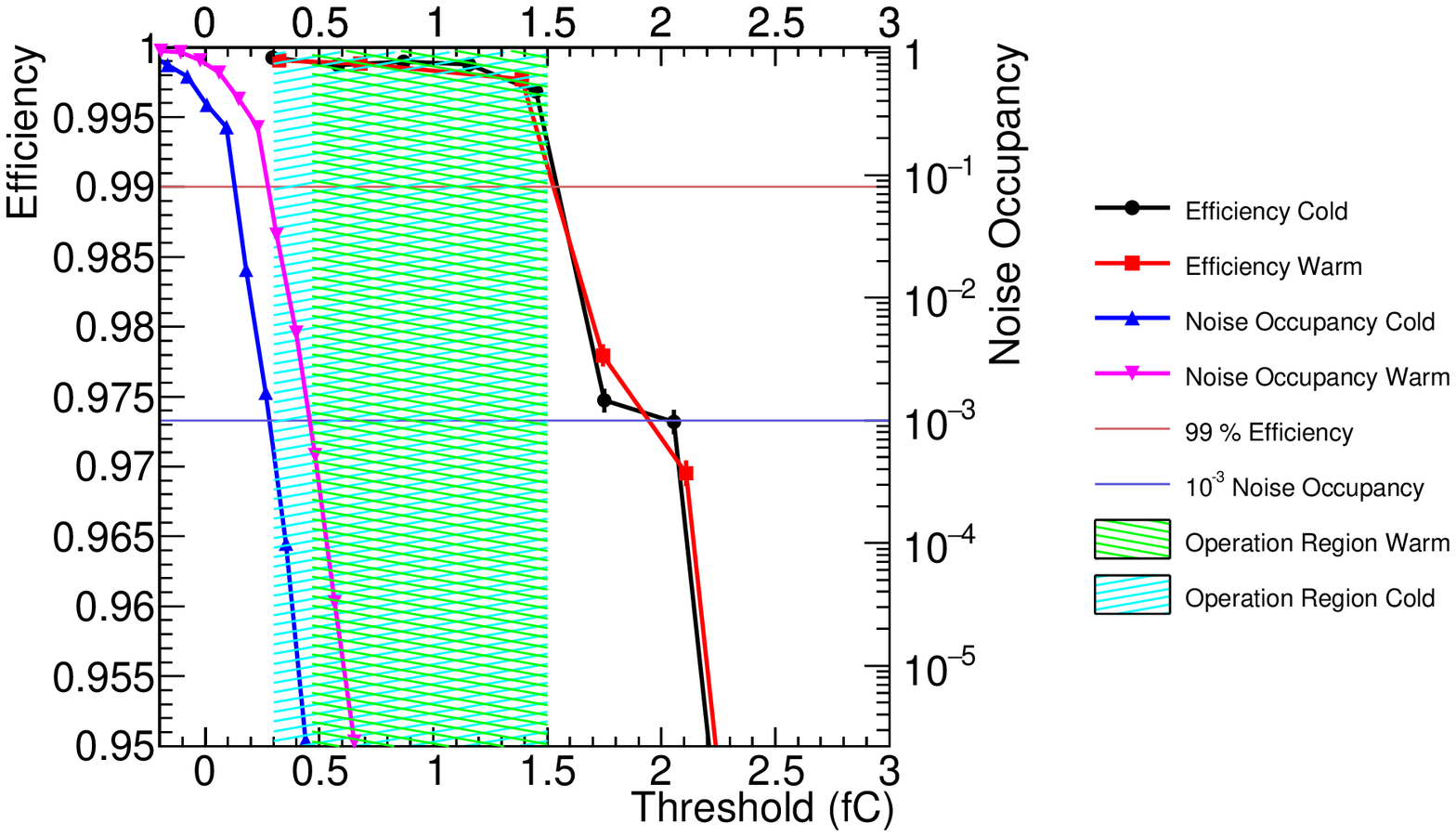}
    \caption{The plot shows the hit efficiency versus threshold (left hand scale) and noise occupancy versus threshold (right hand scale). The shaded areas represent the operational region of the ITk SS module with detection efficiency higher than 99$~\%$ and noise occupancy below~10$^{-3}$ for chip 4 in Hybrid Y, which is representative of the module performance.}
    \label{fig:plot17}
\end{figure}

 To maintain high tracking efficiency the requirement of the ATLAS experiment on the ITk strip detectors is to have the detection efficiency above~99$\%$. At the peak luminosity in the inner barrel layer, the occupancy from "pile-up" (collisions within the same bunch crossing) will be up to $1\%$. In order not to significantly degrade the pattern recognition, the noise occupancy should be at least an order of magnitude smaller than this figure. The specification on the maximum noise occupancy is therefore set to be ~10$^{-3}$. The operational region for the ITk Strip Module defined by the noise occupancy obtained from the pedestal runs and detection efficiency from the tests with the beam is shown in Figure \ref{fig:plot17}. Data for chip 4 in Hybrid Y (as in Figure \ref{fig:plot1} and Figure \ref{fig:plot18}) are used as the beam is centred on this chip.  

\par The signal-to-noise ratio (SNR) is also a critical figure of merit in ITk strip modules. The general requirement for the ATLAS ITk strip modules is to maintain a SNR greater than a value of 10 throughout the lifetime of the ITk~\cite{collaboration2017technical}. 
This is most challenging at the end of lifetime due to the large degradation of  charge signal post-irradiation. There is also a small radiation induced increase in the noise. 
Using the expected degradation in signal and noise from radiation damage~\cite{collaboration2017technical}, the noise for the SS modules before radiation damage should be less than 0.147 fC.  

The SNR values measured for the studied unirradiated ITk SS module are shown in Tab.~\ref{Tab:tab1}.  
The measured mean noise values for the module are less than the upper limit and the signal charge measurement is compatible with expectation. Therefore the expected SNR at the end of lifetime is greater than 10~\cite{collaboration2017technical}.


\begin{table}[h!]
\begin{center}
\begin{tabular}{|c|c|c|c|} 
\hline
 & Mean Noise (fC) & Mean MPV (fC) & SNR  \\
\hline
Cold & 0.0884 $\pm$ 0.005 & 3.19 $\pm$ 0.17 & 36.09 $\pm$ 2.80 \\ 
Warm & 0.1089 $\pm$ 0.002 & 3.18 $\pm$ 0.17 & 29.20 $\pm$ 1.65  \\ 
\hline
\end{tabular}
\caption{Signal-to-noise ratios (SNRs) measured for the ITk SS module when operated at $-30\:\degree\mathrm{C}$ and $+30\:\degree\mathrm{C}$. The estimate of noise and MPV both used the mean values of the distributions. }
\label{Tab:tab1}
\end{center}
\end{table}

\subsubsection{Clustering}
\label{cluster}
\par
\begin{figure}
    \centering
    \includegraphics[width=0.8\textwidth]{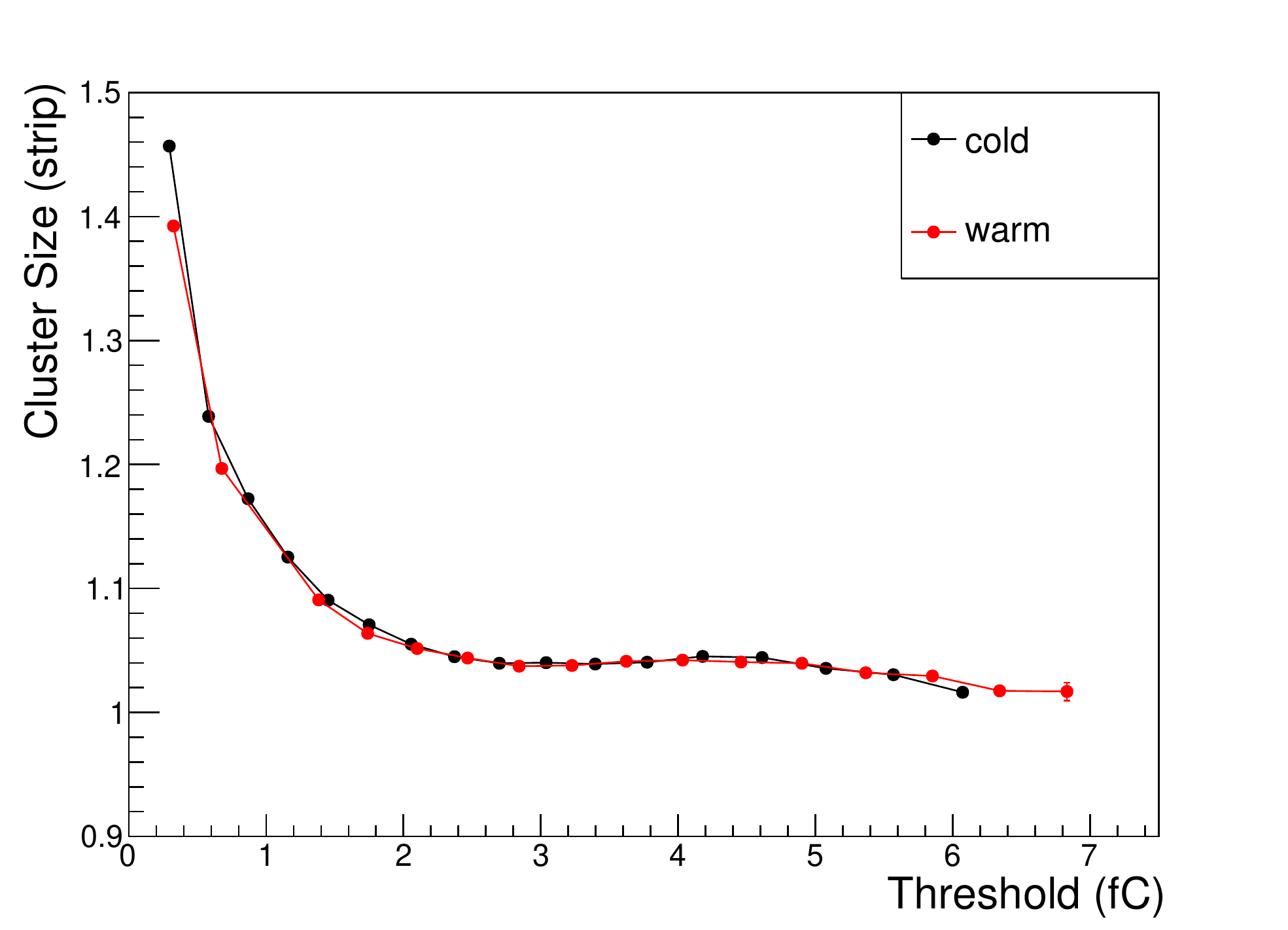}
    \caption{The threshold dependence of the average cluster size measured for the ITk SS module at $-30\:\degree\mathrm{C}$ (black) and at $+30\:\degree\mathrm{C}$ (red). The cluster size cannot go below 1. }
    \label{fig:plot5}
\end{figure}

\begin{figure}
    \centering
    \includegraphics[width=0.8\textwidth]{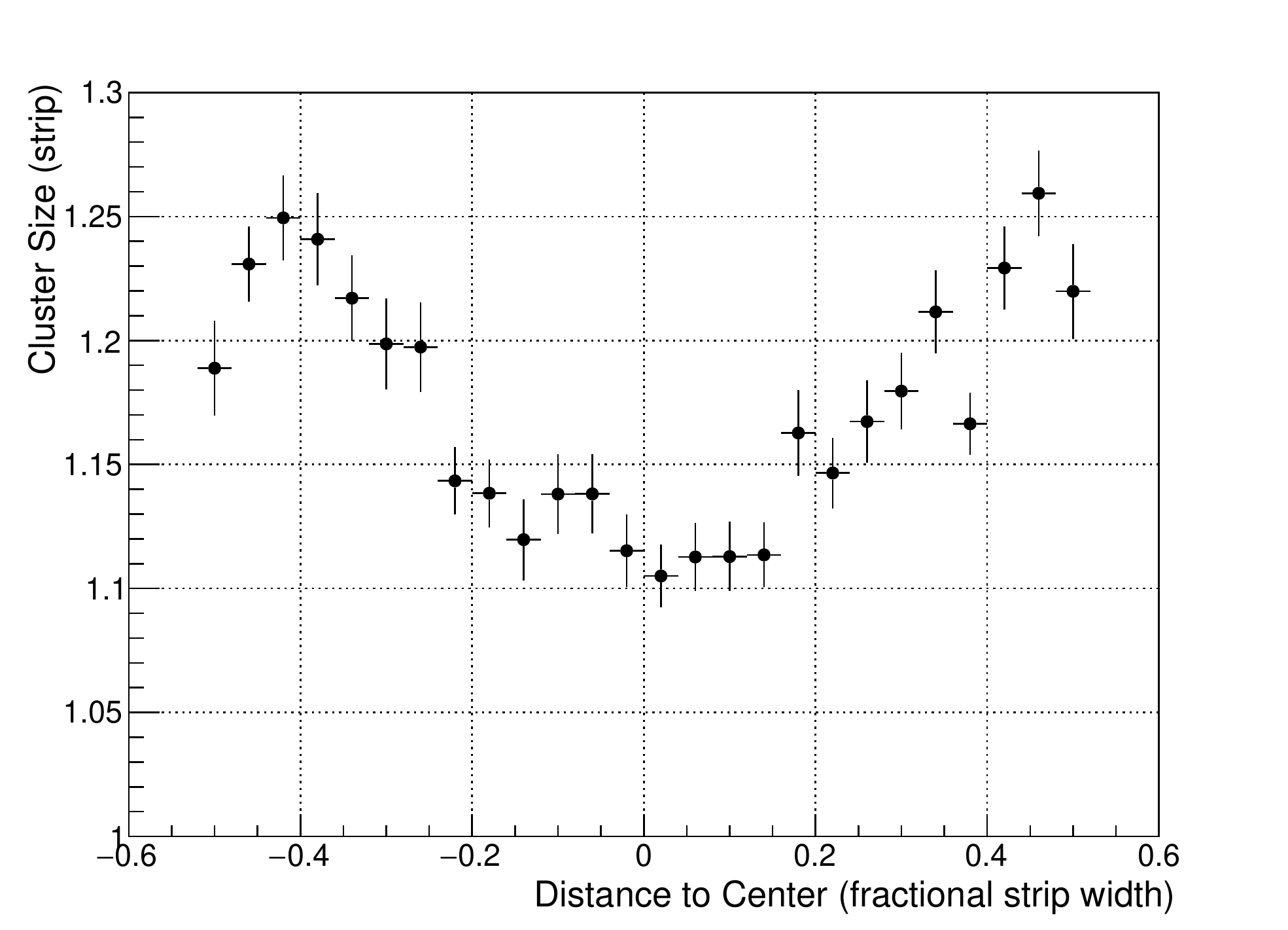}
    \caption{The relation between the cluster size and particle position
when a beam particle traverses a strip. The hit position is in the range
-0.5 to +0.5 (fractional strip width) with 0 being the strip centre. The operational
strip threshold is 0.87 fC.}
    \label{fig:plot6}
\end{figure}

 A large signal on the strip increases the chance for neighbouring strips to detect a signal over the threshold as well.
When the threshold is low, the same particle track is likely to trigger a response on multiple strips. The cluster describes how many consecutive strips are recorded  for the hit of one particle passing through. 
As the threshold is decreased, the system is more sensitive to lower values of charge. Therefore smaller charges on neighbouring strips will be more likely to be above the threshold, thus increasing the cluster size. 
This is demonstrated in Figure~\ref{fig:plot5}. When the particle hits the edge of a strip, the neighbouring strips will see more charge, causing an increase in cluster size. For a single hit cluster the resolution is limited to (strip pitch)/$\sqrt{12}$, whereas higher resolution can be achieved for larger clusters. Figure~\ref{fig:plot6} illustrates the relation between the average cluster size and the position where the particle passed through the strip for the threshold 0.87 fC. 
There is a clearly visible dip in the cluster size in the central region of a strip.

\subsubsection{Pulse Shape Studies}
\label{pulseshape}
\begin{figure}
    \centering
    \includegraphics[width=0.8\textwidth]{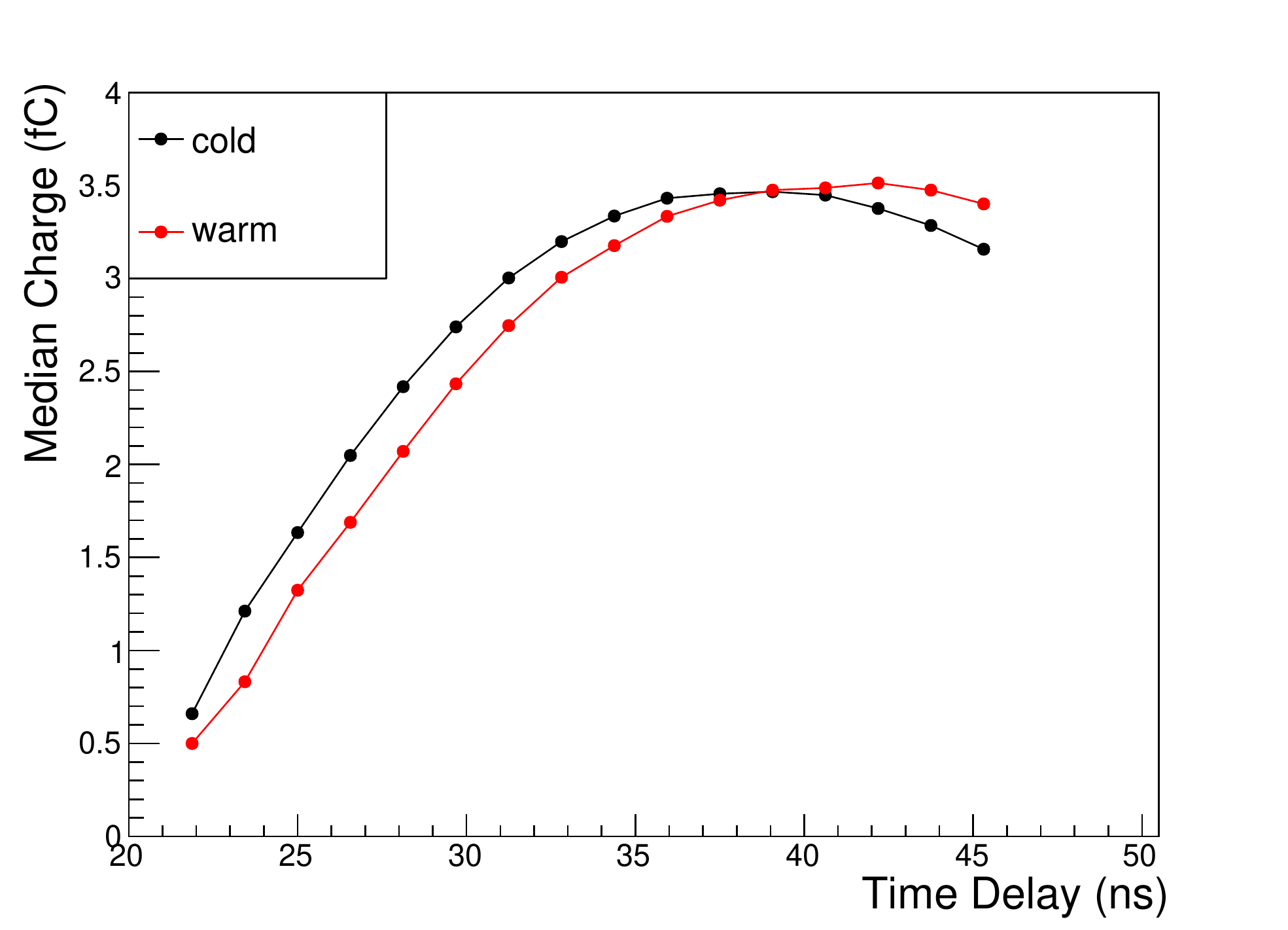}
    \caption{Dependence of the median charge on the selected time delay obtained for the ITk SS module measured at $-30\:\degree\mathrm{C}$ (black) and $+30\:\degree\mathrm{C}$ (red) in September 2019 DESY test beam campaign. The lines drawn between data points are to improve visibility.
    }
    \label{fig:plot8}
\end{figure}

The induced current in the sensor is amplified by a charge sensitive amplifier. After shaping, the pulse lasts for a time corresponding to two LHC bunch crossings. To maximise the signal detected, the pulse should be sampled at the time corresponding to the maximum of the pulse. 
The readout delay is defined as the time difference between the scintillator trigger and the trigger to the DUT. 
The occupancy depends on where the set readout delay falls on the pulse. If the timing window is not chosen correctly, the measured charge does not correspond to the full signal induced by the particle in the silicon sensor. The readout delay setting is used to adjust the selection window for the incoming charge pulse, with 16 time steps available within a \mbox{25 ns} interval. 
For each delay setting, a scan of the hit efficiency of the module as a function of the threshold charge is performed.  This is used to determine the median charge at each delay setting, which is shown in Figure~\ref{fig:plot8} for both the cold and warm measurements. The carrier mobility decreases with increasing temperature which results in the transistor’s transconductance decreasing with temperature. This change in transconductance impacts the bandwidth of pre-amplifier and shaper stages, therefore a faster response is expected at lower temperatures.

\par
\begin{figure}
    \centering
    \includegraphics[width=0.8\textwidth]{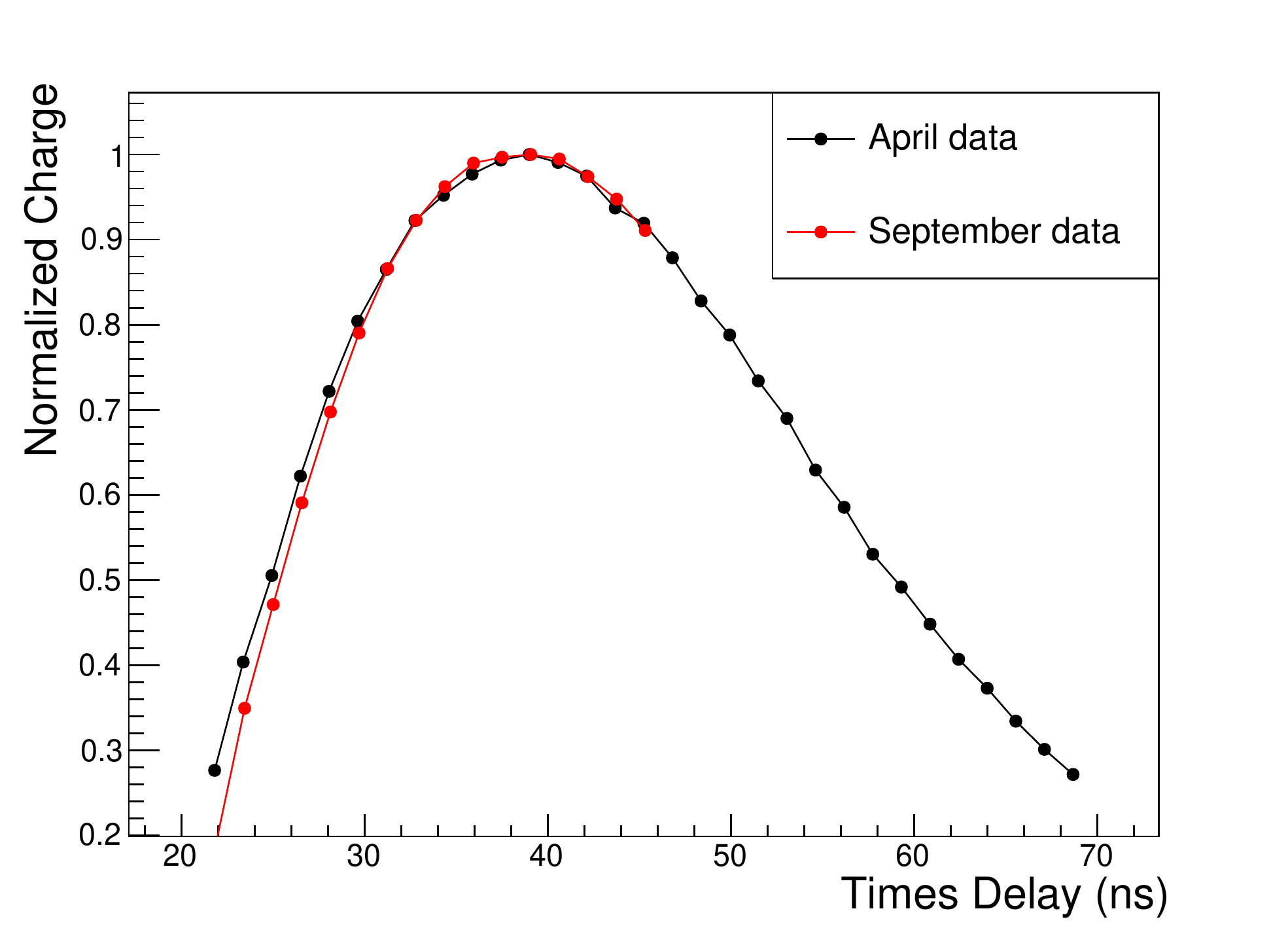}
    \caption{Comparison of the full pulse shape measured for the ITk LS module at the DESY test beam campaign in April 2019 (black)  with the pulse shape obtained for the ITk SS module in September 2019. Both pulse shapes were determined from the warm operation and agree very well with each other.  The offsets in the X axis are adjusted between the two data sets to minimise the differences. The absolute offset has no significance.}
    \label{fig:plot9}
\end{figure}

 
These results give an indication that the pulse is faster at cold than warm temperatures as expected. Due to the limited time available in the experimental area during the
September 2019 test beam campaign, the data do not contain the complete scan of the pulse.
However, the full pulse is present in test beam data taken for the ITk LS module in April 2019.
The April 2019 data were taken with a beam energy of 5.8 GeV. The energy difference compared to the September 2919 data will not result in any significant change in the pulse shape.
The data were taken in "warm" conditions, similar to the warm runs from the September 2019 data.
The stability within a threshold scan was measured using the thermocouple to be $\pm 0.5\:\degree\mathrm{C}$. The estimated uncertainty in the absolute module temperature is $3.0\:\degree\mathrm{C}$.
These temperature uncertainties do not significantly impact the pulse shape.
The pulse shapes measured from warm data in the April 2019 and September 2019 data
are plotted in Figure~\ref{fig:plot9}. It can be seen that September and April 2019 measurements agree with each other in terms of the shape of the rising edge of the pulse.~\footnote{Circuit model calculations show that the effect of the different strip capacitance for SS and LS sensors results in a negligible change in pulse shape.}

\par According to the full pulse shape shown in Figure~\ref{fig:plot9}, the FWHM of the measured data is $34.13 ~\mathrm{ns}$. The height is defined as the maximum point of the curve.  The FWHM is calculated by finding the points where the half-maximum is reached and using interpolation between the neighbouring points to get the width.  The uncertainty of the FWHM comes from the jitter effect of the electronics, the variation in the time duration of each delay bin, and the fact that the pulse shape extracted from test beam data is not continuous. The jitter effect contributes by \mbox{0.15 ns} at most. The variation in the time duration for each delay bin accounts for \mbox{0.2 ns}. Since the data points are discrete, an improved method is introduced to get the FWHM by fitting the pulse curve with a smooth function.  A new FWHM of 34.20 ns was obtained from the smooth curve.  It has a \mbox{0.07 ns} difference from the FWHM obtained from discrete data points.  The \mbox{0.07 ns} difference is used as part of the conservative systematic uncertainty estimation.  The statistical uncertainty is about \mbox{0.09 ns}. These uncertainties combined sum up to \mbox{0.28 ns}. The theoretical FWHM calculated with the circuit model is \mbox{34 ns}~\cite{ref2}. As the pulse from the sensor is 98\% contained within a \mbox{5 ns} bin, the pulse is short enough to be replaced by a delta pulse in circuit modeling on the scale of \mbox{34 ns}. The data agrees with the theoretical calculations within uncertainties.

\FloatBarrier
\section{Conclusions}

The ATLAS ITk strip detector will be operated cold to minimise radiation damage effects so it is vital to understand the change in performance comparing warm and cold operations. A significant decrease in noise when operating the detector cold was observed in agreement with a circuit model of the ABCStar front-end electronics. The magnitude of the reconstructed signal from 5.4 GeV electrons was measured at cold and warm temperatures. No significant difference was observed as expected. The obtained values agree with the GEANT4 simulation. The very large value of the measured SNR resulted in a wide operating window in which the efficiency was above 99\% and the noise occupancy was below $10^{-3}$. These results demonstrate that the SNR will show a very large safety margin at the start of HL-LHC operation. 
The measured noise value at the cold temperature was significantly smaller than the maximum allowed value that would correspond to an acceptable value of the SNR at the end of HL-LHC operation.
The speed of the pulse from the front-end electronics is crucial to minimise the effects of pile-up. Measurements of the magnitude of the pulse as a function of delay were made. These were used to reconstruct the  pulse shape in time. The FWHM of this pulse was in good agreement with the theoretical circuit model calculations. This confirms that the pulse is sufficiently fast for HL-LHC operation.


\acknowledgments

The measurements leading to these results have been performed at the Test Beam Facility at DESY Hamburg (Germany), a member of the Helmholtz Association (HGF). This project has received funding from the European Union’s Horizon 2020 Research and Innovation programme under Grant Agreement no. 654168. Financial support from the U.K. Science and Technology Facilities Council (STFC) is acknowledged. The research was supported and financed in part by the Ministry of Education, Youth and Sports of the Czech Republic coming from the projects LTT17018 Inter-Excellence and LM2018104 CERN-CZ.



\end{document}